\documentclass[fleqn,usenatbib]{mnras}

\usepackage[varg]{newtx}

\usepackage[T1]{fontenc}
\usepackage{bm}
\DeclareRobustCommand{\VAN}[3]{#2}
\let\VANthebibliography\thebibliography
\def\thebibliography{\DeclareRobustCommand{\VAN}[3]{##3}\VANthebibliography}

\usepackage{xcolor}

\usepackage{physics}
\usepackage{graphicx}	
\usepackage{amsmath}	
\usepackage{booktabs}
\usepackage{multirow}
\usepackage{easyReview}
\usepackage{ulem}

\newcommand{\FAC}{\texttt{FAC}}
\newcommand{\GRASP}{\texttt{GRASP2K}}
\newcommand{\HFR}{\texttt{HFR}}
\newcommand{\HULLAC}{\texttt{HULLAC}}

\title[Uranium and Neodymium Opacities]{Opacities of Singly and Doubly Ionised Neodymium and Uranium for Kilonova Emission Modeling}

\author[A.~Flörs et al.]{A.~Fl\"ors,$^{1}$\thanks{E-mail: a.floers@gsi.de},
R.~F.~Silva$^{2,3}$,
J.~Deprince$^{4,5}$,
H.~Carvajal~Gallego$^{5}$,
G.~Leck$^{1,6}$,
L.~J.~Shingles$^{1}$,
\newauthor
G.~Martínez-Pinedo$^{1,6,7}$,
J.~M.~Sampaio $^{2,3}$,
P.~Amaro$^{8}$,
J.~P.~Marques $^{2,3}$,
S.~Goriely $^{4}$,
P.~Quinet $^{5,9}$,
\newauthor
P.~Palmeri$^{5}$,
M.~Godefroid$^{10}$
\\
$^{1}$GSI Helmholtzzentrum f\"ur Schwerionenforschung, Planckstraße 1, 64291 Darmstadt, Germany\\
$^{2}$Laboratório de Instrumentação e Física Experimental de Partículas (LIP), Av. Prof. Gama Pinto 2, 1649-003 Lisboa, Portugal\\
$^{3}$Faculdade de Ciências da Universidade de Lisboa, Rua Ernesto de Vasconcelos, Edifício C8, 1749-016, Lisboa, Portugal\\
$^{4}$Institut d'Astronomie et d'Astrophysique, CP-226, Universit\'e Libre de Bruxelles, B-1050 Brussels, Belgium\\
$^{5}$Physique Atomique et Astrophysique, Universit\'e de Mons, B-7000 Mons, Belgium\\
$^{6}$Institut für Kernphysik (Theoriezentrum), Fachbereich Physik, Technische Universität Darmstadt, Schlossgartenstraße 2, 64289 Darmstadt, Germany\\
$^{7}$Helmholtz Forschungsakademie Hessen für FAIR, GSI Helmholtzzentrum für Schwerionenforschung, Planckstraße 1, 64291 Darmstadt, Germany\\
$^{8}$Laboratory for Instrumentation, Biomedical Engineering and Radiation Physics (LIBPhys-UNL), Department of Physics, NOVA School of Science and Technology, \\ NOVA University Lisbon, 2829-516 Caparica, Portugal\\
$^{9}$ IPNAS, Universit\'e de Li\`ege, B-4000 Li\`ege, Belgium\\
$^{10}$ Spectroscopy, Quantum Chemistry and Atmospheric Remote Sensing, CP160/09, Universit\'e Libre de Bruxelles, B-1050 Brussels, Belgium
}
\date{Accepted XXX. Received YYY; in original form ZZZ}

\pubyear{2022}

\begin{document}
\label{firstpage}
\pagerange{\pageref{firstpage}--\pageref{lastpage}}
\maketitle

\begin{abstract}
Even though the electromagnetic counterpart AT2017gfo to the binary neutron star merger GW170817 is powered by the radioactive decay of r-process nuclei, only few tentative identifications of light r-process elements have been made so far. 
One of the major limitations for the identification of heavy nuclei is incomplete or missing atomic data. While substantial progress has been made on lanthanide atomic data over the last few years, for actinides there has been less emphasis, with the first complete set of opacity data only recently published. We perform atomic structure calculations of neodymium $(Z=60)$ as well as the corresponding actinide uranium $(Z=92)$. Using two different codes (\FAC\,and \HFR) for the calculation of the atomic data, we investigate the accuracy of the calculated data (energy levels and electric dipole transitions) and their effect on kilonova opacities. For the \FAC\,calculations, we optimise the local central potential and the number of included configurations and use a dedicated calibration technique to improve the agreement between theoretical and available experimental atomic energy levels (AELs). For ions with vast amounts of experimental data available, the presented opacities agree quite well with previous estimations. On the other hand, the optimisation and calibration method cannot be used for ions with only few available AELs. For these cases, where no experimental nor benchmarked calculations are available, a large spread in the opacities estimated from the atomic data obtained with the various atomic structure codes is observed.We find that the opacity of uranium is almost double the neodymium opacity.
\end{abstract}




\section{Introduction}
About half of the elements heavier than iron are produced in the astrophysical rapid neutron-capture process (or r-process)~\citep[][]{2021RvMP...93a5002C}. In contrast to the s-process (slow neutron capture process), r-process neutron captures occur much faster than the typical beta-decay timescale of the synthesised nuclei \citep[][]{1957RvMP...29..547B}. One of the most promising r-process production sites are binary neutron star mergers (BNS)~\citep{symbalisty82,Lattimer.Schramm:1974,Eichler.Livio.ea:1989,1999ApJ...525L.121F}, where the r-process is expected to lead to an electromagnetic transient known as kilonova~\citep{Metzger.Martinez.ea:2010}. Such signal was  recently observed following the gravitational wave event GW170817 \citep[][]{2017ApJ...848L..12A, 2017ApJ...848L..13A}.

The observed signal of the kilonova AT2017gfo -- the electromagnetic counterpart following GW170817 -- suggests at least some heavy r-process material was produced \citep[][]{2017Natur.551...80K, 2017PASJ...69..102T, 2017ApJ...850L..37P}. The electromagnetic emission from the first and only spectroscopically observed kilonova was studied extensively in visible and near-infrared (NIR) bands observed by ground-based telescopes for about two weeks until it faded out of reach of the 8-10$\,$m-class facilities \citep[][]{2017Sci...358.1574S, 2017Natur.551...67P, 2017Natur.551...75S, 2017ApJ...848L..17C, 2017Sci...358.1570D, 2017Sci...358.1559K, 2017ApJ...848L..27T}. The tail of the quasi-bolometric light curve of AT2017gfo is in good agreement with the energy release rate from radioactive decays of r-process elements \citep[][]{2017arXiv171005931M}. The rapid spectral evolution from a blue and nearly featureless continuum to a red spectrum (peaking in the NIR), rich in absorption and emission lines as well as the evolutionary timescale of the light curve indicate that high-opacity elements must have been synthesised, with opacities much higher than those typical for the iron group elements (IGE) commonly seen in thermonuclear and core collapse supernovae \citep[][]{2017Natur.551...75S, 2017Natur.551...67P, 2018MNRAS.481.3423W}.

So far, only a single element -- strontium -- has been firmly identified in the kilonova AT2017gfo \citep[][]{2019Natur.574..497W, 2022MNRAS.515..631G}. The location of the proposed Sr feature could also be explained by the \ion{He}{i} $10831$\,\AA\,line. However, to have a noticeable effect on the spectrum, the required helium mass exceeds what is expected to be produced in merger simulations by about one order of magnitude \citep[][]{2022ApJ...925...22P}. In particular, no r-process elements of the second or third peaks have been unambiguously identified \citep[see however][for tentative identifications of \ion{La}{iii} and \ion{Ce}{iii}]{2022ApJ...939....8D}. Unsuccessful searches involve cesium/tellurium \citep[][]{2017Natur.551...75S} and gold/platinum \citep[][]{2021MNRAS.506.3560G}. Identification of spectral features of lanthanides or actinides, together with a constraint on the abundances, would settle the debate on whether BNS mergers are responsible for r-process nuclei seen in the universe, and, if so, whether they are the dominant site of production. A straightforward, albeit very challenging approach to element identification, is through radiative transfer modelling of the observed spectra of AT2017gfo, as was done with strontium \citep[][]{2019Natur.574..497W}. 

Radiative transfer models in their simplest form (1D, local thermodynamic equilibrium [LTE], no time dependence) still require precise knowledge of level energies and bound-bound atomic transitions, which make up the bulk of the photon opacity in the r-process enriched ejecta. It is expected that lanthanide and actinide ions each have of order 10$^6$ relevant transitions -- a factor of 10--100 more than IGE ions \citep[][]{2013ApJ...774...25K} -- while only a tiny fraction has been measured for a few selected ions \citep[see e.g. experimental data on the NIST ASD;][]{NIST_ASD}. The ejecta contain only neutral to $\approx 4$ times ionised atoms at phases beyond 1\,day.  It is not experimentally feasible to measure a full set of opacities for all quasi-neutral ions from the IGE to the actinides ($\approx 250$ relevant ions), and thus the only way of obtaining complete atomic data is through theoretical atomic structure calculations. Since the observation of AT2017gfo, several calculations of weakly ionised r-process opacities have been published \citep[][]{2019ApJS..240...29G, 2022MNRAS.517..281G, 2020ApJS..248...13G, 2020ApJS..248...17R, 2021MNRAS.506.3560G, 2020MNRAS.493.4143F, 2020MNRAS.496.1369T, 2022MNRAS.510.3806P, 2022Atoms..10...18S, 2021MNRAS.501.1440C}, focusing mainly on lanthanides. However, if material with sufficiently low electron fraction $Y_e$ ($\approx 0.15$ or lower) is ejected in the merging process, nucleosynthesis can proceed to the actinides, which are expected to have photon opacities similar or even higher than lanthanides. For actinides, only few atomic structure calculations have been performed \citep{2020ApJ...899...24E, 2023MNRAS.519.2862F}.

In this paper we present new calculations of neodymium and uranium, which act as case studies of the lanthanides and actinides. We focus on the singly and doubly ionised ions due to their importance in the line-forming regions, based on kilonova models produced with radiative transfer codes. To investigate variations in the calculated atomic properties we use two atomic structure codes: the \texttt{Flexible Atomic Code} \citep[\FAC,][]{2008CaJPh..86..675G} and the \texttt{Hartree-Fock-Relativistic} code \citep[\HFR,][]{1981tass.book.....C}, described in Sections~\ref{sec:fac} and \ref{sec:hfr}, respectively. While we compute \textit{ab-initio} atomic data with the \HFR\,code, we calibrate the local central potential and the calculated level energies in the \FAC\,calculations to experimental data (Sections~\ref{sec:Nd_atomic_data} and \ref{sec:U_atomic_data}). Finally, in Section~\ref{sec:opacities} we discuss and compare the resulting atomic opacities to published data within the expansion opacity \citep[data from ][]{2019ApJS..240...29G, 2020MNRAS.496.1369T}, line binned opacity \citep[data from][]{2020MNRAS.493.4143F, 2023MNRAS.519.2862F, nist_opac}, and Planck mean opacity frameworks. Conclusions are drawn in Section~\ref{sec:conclusions}.

\section{Methods}

\subsection{FAC calculations}
\label{sec:fac}
Part of the calculations for this work were performed using the open source and freely available \texttt{Flexible Atomic Code}\,(\FAC) \citep{2008CaJPh..86..675G} relativistic atomic structure package based on the diagonalization of the Dirac-Coulomb Hamiltonian given, in atomic units, as
\begin{equation}
    H_{D C}=\sum_{i=1}^{N}\left(c \bm{\alpha}_{i} \cdot \bm{p}_{i}+\left(\beta_{i}-1\right) c^{2}+V_{i}\right)+\sum_{i<i}^{N} \frac{1}{r_{i j}} \, , 
\end{equation}
where $\bm{\alpha}_{i}$ and $\beta_i$ are the $ 4 \, \times \, 4$ Dirac matrices and $V_i$ accounts for potential due to the nuclear charge. Recoil and retardation effects are included in the Breit interaction in the zero-energy limit for the exchanged photon, whereas vacuum polarisation and self-energy corrections are treated in the screened hydrogenic approximation. 

Relativistic configuration interaction (CI) calculations are performed based on a set of basis states, configuration state functions (CSFs), which consist of linear combinations of antisymmetrised products of $N$ one-electron Dirac spinors,
\begin{equation}
    \varphi_{n \kappa m}=\left(\begin{array}{c}
                         P_{n \kappa}(r) \chi_{\kappa m}(\theta, \phi, \sigma) \\
                         i Q_{n \kappa}(r) \chi_{-\kappa m}(\theta, \phi, \sigma)
                         \end{array}\right),
\end{equation}
where $\chi_{\kappa m}$ represents the spin-angular function and $P_{n \kappa}$ and $Q_{n \kappa}$ are the radial functions of the large and small components, respectively. Successive shells are coupled using a $jj$ coupling scheme. Atomic state functions (ASFs) $\Psi$ are then constructed from a superposition of $i=1,\cdots , N_{\texttt{CSF}}$ configuration state functions (CSFs) $\varphi_i$ with the same $J$ and parity $P$ symmetry, 
\begin{equation}
    \Psi (\gamma  J M_J P)=\sum_i^{N_{\texttt{CSF}}}c_i \; \varphi_i ( \gamma_i  J M_J P), 
\end{equation}
where the $\gamma_i$ stands for the complete relevant information to define each CSF (configuration and coupling tree  quantum numbers). \\

The mixing coefficients $\{c_i\}$ are obtained by solving the eigenvalue problem ${\mathbf H} {\bf c} = E {\mathbf c} $, with ${\mathbf c} = (c_1, c_2, \ldots , c_{N_{\texttt{CSF}}} )^t $. The eigenvalues obtained from the diagonalization of the Hamiltonian matrix ${\mathbf H}$ are, therefore, the best approximation for the energies in the space described by the basis of the selected CSFs. Increasing the number of CSFs used would improve the wave functions, and hence, the expected accuracy of the AELs, but the improvement is not expected to be significant enough to outweigh the increasing computational cost. Therefore, we search for the optimal set of CSFs by examining how the level energies converge as the number of configurations increases.

\FAC\, uses a variant of the conventional Dirac-Fock-Slater method to compute one-electron radial functions. The small and large components are determined by solving self-consistently the coupled Dirac equation for a local central potential $V(r)$

\begin{equation}
    \begin{array}{l}
        \displaystyle{\left(\frac{d}{d r}+\frac{\kappa}{r}\right) P_{n \kappa} (r)=\alpha\left(\varepsilon_{n \kappa}-V(r)+\frac{2}{\alpha^{2}}\right) Q_{n \kappa} (r)} \\[5mm]
        \displaystyle{\left(\frac{d}{d r}-\frac{\kappa}{r}\right) Q_{n \kappa} (r) =\alpha\left(-\varepsilon_{n \kappa}+V(r) \right) P_{n \kappa} (r),}
    \end{array}
    \label{dirac equations}
\end{equation}
where $\alpha$ is the fine structure constant and $\varepsilon_{n \kappa}$ are the one-electron orbital energies. Although not as accurate as the multiconfiguration Dirac-Fock method used in the \GRASP\, \citep{2013CoPhC.184.2197J} and \texttt{MCDFGME} \citep{1975CoPhC...9...31D,1995PhRvA..51.1132I} structure codes, this approach has many advantages. As the same potential is felt by all the electrons of the system, all orbitals are automatically orthogonal. Moreover, the secular equation necessary to determine the eigenvalues and the mixing coefficients has to be solved only once, making the calculations much faster and computationally efficient when compared to the other codes mentioned. \FAC\, has been shown to provide particularly good results when compared to experiments and other structure codes for calculations on lighter and/or highly excited ions \citep{2008CaJPh..86..675G}.

Following a similar approach to the one first described by \citet{1989PhRvA..40..604S} and \citet{1989PhRvA..40..616Z}, a single fictitious mean configuration (FMC) with fractional occupation numbers is adopted. The local central potential is then derived from this mean configuration using a self-consistent Dirac-Fock-Slater iteration. This unique potential is then used in Equation \ref{dirac equations} in order to determine the remaining one-electron radial orbitals. Although this reduces the computation time and avoids convergence issues, as the orthogonality of the different orbitals with the same $\kappa$-value is automatically ensured with this method, the potential is not optimised for a single configuration. 
To accommodate multiple configurations, the FMC approach offers a good compromise for reducing the overall error of the calculation and getting a satisfactory accuracy of individual level energies. 
Typically, for the construction of the FMC, the occupation of the active electrons is split equally between a set of configurations. However, the weight of each configuration can be changed in order to raise (or reduce) its contribution in the construction of the mean configuration.

A summary of the calculations achieved, for each ion, with \FAC\, is given in Table \ref{tab:FAC_configs}. Multiple models were computed, with an increasing number of configurations in order to test the sensitivity of the level energy to a higher degree of correlation and of the opacity with the inclusion of higher lying states.
The configurations were sorted by the lowest level's excitation energy and then successively included in the various models considered. The largest model considered included 30 configurations, for which they were used in relativistic configuration interaction calculations; however, due to computational limits, only oscillator strengths and wavelengths of radiative transitions between the lowest 22 configurations were included. \\

\begin{table*}
	\caption{Configurations included in the \FAC\, calculations for the different models tested. The presented configurations are sorted by energy. Boldface configurations are used for the radial optimisation of the potential. The weights associated with each configuration are also shown, ordered in the same way as the boldface configurations in the \textit{Configurations} column. Consecutive rows for the same ion show the additional configurations included with respect to the base model.}
	\label{tab:FAC_configs}
	\begin{tabular}{@{}cccl@{}}
		\toprule
		Ion & FMC weights & Model & Configurations \\ 
		\midrule
		\multirow{4}{*}{Nd II} &
		\multirow{4}{*}{(0.1, 0.1, 0.2, 0.1, 0.8)}&
		-- 8 conf. &
		\begin{tabular}{l} $\bm{4\textit{\textbf{f}}^{\,4}\,6\textit{\textbf{s}}^{\,1}}$, $\bm{4\textit{\textbf{f}}^{\,3}\,5\textit{\textbf{d}}^{\,2}}$, $\bm{4\textit{\textbf{f}}^{\,4}\,5\textit{\textbf{d}}^{\,1}}$, 4\textit{f}$^{3}$\,5\textit{d}$^{1}$ 6\textit{s}$^{1}$, 4\textit{f}$^{3}$\,6\textit{s}$^{2}$, 4\textit{f}$^{4}$\,6\textit{p}$^{1}$, $\bm{4\textit{\textbf{f}}^{\,3}\,5\textit{\textbf{d}}^{\,1}\,6\textit{\textbf{p}}^{\,1}}$, $\bm{4\textit{\textbf{f}}^{\,3}\,6\textit{\textbf{s}}^{\,1}\,6\textit{\textbf{p}}^{\,1}}$\\\end{tabular} \\
		& & 
		-- 15 conf. &
		\begin{tabular}{l} 4\textit{f}$^{4}$\,7\textit{s}$^{1}$, 4\textit{f}$^{4}$\,6\textit{d}$^{1}$, 4\textit{f}$^{4}$\,7\textit{p}$^{1}$, 4\textit{f}$^{4}$\,5\textit{f}$^{1}$, 4\textit{f}$^{4}$\,8\textit{s}$^{1}$, 4\textit{f}$^{4}$\,7\textit{d}$^{1}$, 4\textit{f}$^{4}$\,8\textit{p}$^{1}$ \\\end{tabular} \\
		& & 
		-- 22 conf. &
		\begin{tabular}{l}4\textit{f}$^{4}$\,6\textit{f}$^{1}$, 4\textit{f}$^{4}$\,5\textit{g}$^{1}$, 4\textit{f}$^{3}$\,5\textit{d}$^{1}$\,6\textit{d}$^{1}$,  4\textit{f}$^{4}$\,8\textit{d}$^{1}$, 4\textit{f}$^{4}$\,7\textit{f}$^{1}$, 4\textit{f}$^{4}$\,6\textit{g}$^{1}$, 4\textit{f}$^{4}$\,8\textit{f}$^{1}$\\ \end{tabular} \\
		& & 
		-- extra CI &
		\begin{tabular}{l} 4\textit{f}$^{4}$\,7\textit{g}$^{1}$, 4\textit{f}$^{3}$\,5\textit{d}$^{1}$\,5\textit{f}$^{1}$, 4\textit{f}$^{4}$\,8\textit{f}$^{1}$, 4\textit{f}$^{3}$\,6\textit{p}$^{2}$, 4\textit{f}$^{3}$\,6\textit{s}$^{1}$\,6\textit{d}$^{1}$, 4\textit{f}$^{3}$\,6\textit{s}$^{1}$\,7\textit{s}$^{1}$, 4\textit{f}$^{3}$\,5\textit{d}$^{1}$\,6\textit{f}$^{1}$, 4\textit{f}$^{3}$\,5\textit{d}$^{1}$\,5\textit{g}$^{1}$\end{tabular} \\ \midrule
		\multirow{4}{*}{Nd III} &
		\multirow{4}{*}{(0.4, 0.3, 0.3)}&
		-- 8 conf. & \begin{tabular}{l} $\bm{4\textit{\textbf{f}}^{\,4}}$, $\bm{4\textit{\textbf{f}}^{\,3}\,5\textit{\textbf{d}}^{\,1}}$, $\bm{4\textit{\textbf{f}}^{\,3}\,6\textit{\textbf{s}}^{\,1}}$, 4\textit{f}$^{3}$\,6\textit{p}$^{1}$, 4\textit{f}$^{3}$\,6\textit{d}$^{1}$, 4\textit{f}$^{3}$\,5\textit{f}$^{1}$, 4\textit{f}$^{3}$\,7\textit{s}$^{1}$, 4\textit{f}$^{3}$\,7\textit{p}$^{1}$\end{tabular}   \\  
		& &
		-- 15 conf. & \begin{tabular}{l} 4\textit{f}$^{2}$\,5\textit{d}$^{2}$, 4\textit{f}$^{3}$\,6\textit{f}$^{1}$, 4\textit{f}$^{3}$\,5\textit{g}$^{1}$, 4\textit{f}$^{3}$\,8\textit{s}$^{1}$, 4\textit{f}$^{3}$\,7\textit{d}$^{1}$, 4\textit{f}$^{3}$\,6\textit{g}$^{1}$, 4\textit{f}$^{3}$\,8\textit{p}$^{1}$ \end{tabular}   \\  
		& & 
		-- 22 conf. & \begin{tabular}{l} 4\textit{f}$^{3}$\,8\textit{d}$^{1}$, 4\textit{f}$^{3}$\,7\textit{f}$^{1}$, 4\textit{f}$^{2}$\,5\textit{d}$^{1}$\,6\textit{s}$^{1}$, 4\textit{f}$^{3}$\,8\textit{f}$^{1}$, 4\textit{f}$^{3}$\,7\textit{g}$^{1}$, 4\textit{f}$^{3}$\,8\textit{g}$^{1}$, 4\textit{f}$^{2}$\,5\textit{d}$^{1}$\,6\textit{p}$^{1}$\end{tabular}   \\  
		& &
		-- extra CI & \begin{tabular}{l} 4\textit{f}$^{2}$\,6\textit{s}$^{2}$, 4\textit{f}$^{2}$\,6\textit{s}$^{1}$\,6\textit{p}$^{1}$, 4\textit{f}$^{2}$ 5\textit{d}$^{1}$\,6\textit{d}$^{1}$, 4\textit{f}$^{2}$\,5\textit{d}$^{1}$\,5\textit{f}$^{1}$, 4\textit{f}$^{2}$\,5\textit{d}$^{1}$\,6\textit{f}$^{1}$, 4\textit{f}$^{2}$\,6\textit{p}$^{2}$, 4\textit{f}$^{2}$\,5\textit{d}$^{1}$\,5\textit{g}$^{1}$, 4\textit{f}$^{2}$\,5\textit{d}$^{1}$\,6\textit{g}$^{1}$\end{tabular} \\ \midrule
		\multirow{4}{*}{U II} &
		\multirow{4}{*}{(1.0, 0.95)}&
		-- 8 conf. & \begin{tabular}{l} $\bm{5\textit{\textbf{f}}^{\,3}\,7\textit{\textbf{s}}^{\,2}}$, 5\textit{f}$^{3}$\,6\textit{d}$^{1}$\,7\textit{s}$^{1}$, 5\textit{f}$^{3}$\,6\textit{d}$^{2}$, 5\textit{f}$^{4}$\,7\textit{p}$^{1}$, 5\textit{f}$^{4}$\,7\textit{s}$^{1}$, 5\textit{f}$^{4}$\,6\textit{d}$^{1}$, 5\textit{f}$^{3}$\,6\textit{d}$^{1}$\,7\textit{p}$^{1}$, $\bm{5\textit{\textbf{f}}^{\,3}\,7\textit{\textbf{s}}^{\,1}\,7\textit{\textbf{p}}^{\,1}}$\end{tabular}   \\ 
		& &
		-- 15 conf. & \begin{tabular}{l} 5\textit{f}$^{3}$\,7\textit{s}$^{1}$\,9\textit{s}$^{1}$, 5\textit{f}$^{4}$\,7\textit{g}$^{1}$, 5\textit{f}$^{3}$\,7\textit{s}$^{1}$\,8\textit{d}$^{1}$, 5\textit{f}$^{4}$\,8\textit{g}$^{1}$, 5\textit{f}$^{3}$\,7\textit{s}$^{1}$\,9\textit{p}$^{1}$, 5\textit{f}$^{4}$\,9\textit{g}$^{1}$, 5\textit{f}$^{3}$\,7\textit{s}$^{1}$\,7\textit{f}$^{1}$\end{tabular}   \\ 
		& &
		-- 22 conf. & \begin{tabular}{l} 5\textit{f}$^{2}$\,7\textit{s}$^{2}$\,7\textit{p}$^{1}$, 5\textit{f}$^{3}$\,5\textit{g}$^{1}$\,7\textit{s}$^{1}$, 5\textit{f}$^{3}$\,7\textit{s}$^{1}$\,9\textit{d}$^{1}$, 5\textit{f}$^{3}$\,7\textit{s}$^{1}$\,8\textit{f}$^{1}$, 5\textit{f}$^{3}$\,6\textit{g}$^{1}$\,7\textit{s}$^{1}$, 5\textit{f}$^{3}$\,7\textit{s}$^{1}$\,9\textit{f}$^{1}$, 5\textit{f}$^{3}$\,7\textit{s}$^{1}$\,7\textit{g}$^{1}$\end{tabular}   \\ 
		& &
		-- extra CI & \begin{tabular}{l} 5\textit{f}$^{3}$\,7\textit{p}$^{1}$\,7\textit{d}$^{1}$, 5\textit{f}$^{3}$\,7\textit{s}$^{1}$\,8\textit{g}$^{1}$, 5\textit{f}$^{3}$\,7\textit{s}$^{1}$\,9\textit{g}$^{1}$, 5\textit{f}$^{2}$\,6\textit{d}$^{1}$\,7\textit{s}$^{1}$\,7\textit{d}$^{1}$, 5\textit{f}$^{2}$\,7\textit{s}$^{1}$\,7\textit{p}$^{2}$, 5\textit{f}$^{2}$\,7\textit{s}$^{2}$\,7\textit{d}$^{1}$, 5\textit{f}$^{3}$\,7\textit{d}$^{2}$, 5\textit{f}$^{2}$\,7\textit{s}$^{1}$\,7\textit{p}$^{1}$\,7\textit{d}$^{1}$\end{tabular}   \\ \midrule
		\multirow{4}{*}{U III} &
		\multirow{4}{*}{(0.01, 4.0, 9.0, 0.5)}&
		-- 8 conf. & \begin{tabular}{l} $\bm{5\textit{\textbf{f}}^{\,4}}$, $\bm{5\textit{\textbf{f}}^{\,3}\,6\textit{\textbf{d}}^{\,1}}$, $\bm{5\textit{\textbf{f}}^{\,3}\,7\textit{\textbf{s}}^{\,1}}$, 5\textit{f}$^{3}$\,7\textit{p}$^{1}$, 5\textit{f}$^{2}$\,6\textit{d}$^{2}$, 5\textit{f}$^{2}$\,6\textit{d}$^{1}$\,7\textit{s}$^{1}$, $\bm{5\textit{\textbf{f}}^{\,3}\,7\textit{\textbf{d}}^{\,1}}$, 5\textit{f}$^{3}$\,6\textit{f}$^{1}$ \end{tabular}   \\ 
		& &
		-- 15 conf. & \begin{tabular}{l} 5\textit{f}$^{3}$\,8\textit{s}$^{1}$, 5\textit{f}$^{2}$\,6\textit{d}$^{1}$\,7\textit{p}$^{1}$, 5\textit{f}$^{3}$\,8\textit{p}$^{1}$, 5\textit{f}$^{3}$\,5\textit{g}$^{1}$, 5\textit{f}$^{3}$\,7\textit{f}$^{1}$, 5\textit{f}$^{2}$\,7\textit{s}$^{1}$\,7\textit{p}$^{1}$, 5\textit{f}$^{3}$\,6\textit{g}$^{1}$\end{tabular}   \\ 
		& &
		-- 22 conf. & \begin{tabular}{l} 5\textit{f}$^{3}$\,9\textit{s}$^{1}$, 5\textit{f}$^{3}$\,8\textit{d}$^{1}$, 5\textit{f}$^{3}$\,9\textit{p}$^{1}$, 5\textit{f}$^{3}$\,7\textit{g}$^{1}$, 5\textit{f}$^{3}$\,9\textit{d}$^{1}$, 5\textit{f}$^{3}$\,8\textit{f}$^{1}$, 5\textit{f}$^{3}$\,9\textit{f}$^{1}$ \end{tabular}   \\ 
		& &
		-- extra CI & \begin{tabular}{l} 5\textit{f}$^{2}$\,6\textit{d}$^{1}$\,7\textit{d}$^{1}$, 5\textit{f}$^{2}$\,6\textit{d}$^{1}$\,6\textit{f}$^{1}$, 5\textit{f}$^{3}$\,8\textit{g}$^{1}$, 5\textit{f}$^{3}$\,9\textit{g}$^{1}$, 5\textit{f}$^{2}$\,7\textit{p}$^{2}$, 5\textit{f}$^{2}$\,7\textit{s}$^{1}$\,7\textit{d}$^{1}$, 5\textit{f}$^{2}$\,6\textit{f}$^{1}$\,7\textit{s}$^{1}$, 5\textit{f}$^{2}$\,6\textit{d}$^{1}$\,7\textit{f}$^{1}$\end{tabular}   \\ \bottomrule
	\end{tabular}
\end{table*}

In all calculations using \FAC, the potential was adjusted to reduce the difference between our computed values and the published data in the Atomic Spectra Database (ASD) of the National Institute of Standards and Technology (NIST) database \citep{NIST_ASD, 2022Atoms..10...42K} for the case of Nd \citep{1978aelr.book.....M}, while for U the Selected Constants Energy Levels and Atomic Spectra of Actinides (from now on abbreviated as SCASA), available as an online database \citep[see][]{SCASA}, was used. This was done by manually changing the contribution of the configurations used in the construction of the FMC in order to better reproduce the experimental energy of the lowest levels for each $J$ and parity values. For each ion, roughly 100 values were tested before choosing the one that best matched the experimental data available. The final set of weights included in the calculations is also presented in Table \ref{tab:FAC_configs}. 

Semi-empirical corrections to the energies can be added subsequently to correct for errors induced by the use of a mean configuration. However, when this FAC functionality is applied,  we have noticed significant disparities between the energy levels computed and the reported experimental NIST values. Similar findings have also been reported in \citet{2021PhRvA.103b2808L} and \citet{2022MNRAS.509.4723M}. As a final step, when possible, we use a calibration technique that moves all the levels of a given symmetry by the same energy shift, estimated by the difference of the calculated and measured excitation energies of the lowest level of the considered ($J$-$P$) block. That calibration process, used for all computed ions, does not affect neither the orbitals, nor the ASF wave function compositions, but corrects the transition data through the transition energies and the partition functions through the excitation energies.

\subsection{HFR calculations}
\label{sec:hfr}

The \texttt{Hartree-Fock Relativistic} code (\HFR) was developed by \citet{1981tass.book.....C}. In this computational approach, a set of orbitals is obtained for each configuration by solving the Hartree-Fock (HF) equations, which arise from a variational principle applied to the configuration average energy. Some relativistic corrections are also included in a perturbative way, namely the Blume-Watson spin-orbit (including the one-body Breit interaction operator), mass-variation and one-body Darwin terms. The self-consistent field method is used to solve the coupled HF equations.

In the Slater-Condon approach, the atomic wavefunctions (eigenfunctions of the Hamiltonian) are built as a superposition of basis wavefunctions in the $LSJ\pi$ representation, \textit{i.e.} 

\begin{equation}
    \label{eq:CSF_exp_HFR}
    \Psi(\gamma J M_J P)= \sum_i^{N_{\texttt{CSF}}}c_i \; \phi(\gamma_i L_i S_iJ M_J P).
\end{equation}
The construction and diagonalization of the multiconfiguration Hamiltonian matrix is carried out within the framework of the Slater-Condon theory. Each matrix element is computed as a sum of products of Racah angular coefficients and radial Slater and spin-orbit integrals: 
\begin{equation}
    \mel{i}{H}{j} = \sum_l v^l_{ij} x_l.
\end{equation}
In the present computations, scaling factors of 0.85 are applied to the Slater integrals, as recommended by \citet{1981tass.book.....C}. It was recently shown that the choice of scaling factors between 0.8 and 0.95 virtually does not affect the computed expansion opacities \citep{2023MNRAS.518..332C}. 

The eigenvalues and eigenstates obtained in this way can then be used to compute the radiative wavelengths and oscillator strengths for each possible transition. \\

All the configurations included in the multiple models used in our \HFR\, computations are listed in Table \ref{tab:HFR_configs}. Unlike the strategy followed with \FAC, a more conventional approach was used for our \HFR\, calculations, in which single and double electron substitutions from reference configurations were included. In the case of \ion{U}{ii} and \ion{U}{iii}, several models were considered and tested, with an increasing number of configurations obtained by considering single and/or double electron excitations from reference configurations to higher orbitals with an increasing principal quantum number $n$ (each model includes the same configurations as the previous one in addition to new configurations). The configurations were chosen by considering, in the first models (named $n=6$), the ground configurations (5\textit{f}$^3$7\textit{s}$^2$ and 5\textit{f}$^4$ for \ion{U}{ii} and \ion{U}{iii}, respectively) and single excitations from the reference configurations 5\textit{f}$^5$ (\ion{U}{ii}) and 5\textit{f}$^4$ (\ion{U}{iii}) to all the $n=6$ orbitals. In the second model ($n=7$), single excitations from the same reference configurations to all $n=7$ orbitals were added, as well as configurations arising from a few number of double excitations to 6\textit{d}, 7\textit{s}, 7\textit{p} and 7\textit{d} subshell, in addition to the configurations from the $n=6$ model. In the last two models ($n=8$ and $n=9$), configurations obtained by considering single excitations from the reference configurations to $n=8$ and $n=9$ orbitals, respectively, were successively added.

The motivation of the several models tested in our \HFR\, computations was to assess the convergence of the expansion opacities obtained when using the \HFR\, atomic data coming from the various multiconfiguration models, which will be discussed in Section \ref{sec:opacities}.

\begin{table*}
	\caption{Configurations included in the \HFR\, calculations for the different ions and models tested.}
	\label{tab:HFR_configs}
	\begin{tabular}{@{}ccl@{}}
		\toprule
		Ion & Model    & Configurations \\ \midrule
		$\ion{Nd}{ii}$ &
		-- $n=8$ &
		\begin{tabular}{l} 4\textit{f}$^{5}$, 4\textit{f}$^{4}$\,5\textit{d}$^{1}$, 4\textit{f}$^{4}$\,5\textit{f}$^{1}$, 4\textit{f}$^{4}$\,5\textit{g}$^{1}$, 4\textit{f}$^{4}$\,6\textit{s}$^{1}$, 4\textit{f}$^{4}$\,6\textit{p}$^{1}$, 4\textit{f}$^{4}$\,6\textit{d}$^{1}$, 4\textit{f}$^{4}$\,6\textit{f}$^{1}$, 4\textit{f}$^{4}$\,6\textit{g}$^{1}$, 4\textit{f}$^{4}$\,7\textit{s}$^{1}$, 4\textit{f}$^{4}$\,7\textit{p}$^{1}$, 4\textit{f}$^{4}$\,7\textit{d}$^{1}$, 4\textit{f}$^{4}$\,7\textit{f}$^{1}$, 4\textit{f}$^{4}$\,7\textit{g}$^{1}$, 4\textit{f}$^{4}$\,8\textit{s}$^{1}$, \\4\textit{f}$^{4}$\,8\textit{p}$^{1}$, 4\textit{f}$^{4}$\,8\textit{d}$^{1}$, 4\textit{f}$^{4}$\,8\textit{f}$^{1}$, 4\textit{f}$^{4}$\,8\textit{g}$^{1}$, 4\textit{f}$^{3}$\,5\textit{d}$^{2}$, 4\textit{f}$^{3}$\,5\textit{d}$^{1}$\,6\textit{s}$^{1}$, 4\textit{f}$^{3}$\,5\textit{d}$^{1}$\,6\textit{p}$^{1}$, 4\textit{f}$^{3}$\,5\textit{d}$^{1}$\,6\textit{d}$^{1}$, 4\textit{f}$^{3}$\,6\textit{s}$^{2}$, 4\textit{f}$^{3}$\,6\textit{s}$^{1}$\,6\textit{p}$^{1}$, 4\textit{f}$^{3}$\,6\textit{s}$^{1}$\,6\textit{d}$^{1}$ \\\end{tabular} \\\midrule
		         
		$\ion{Nd}{iii}$ &
		-- $n=8$ & \begin{tabular}{l} 4\textit{f}$^4$, 4\textit{f}$^3$\,5\textit{d}$^{1}$, 4\textit{f}$^3$\,5\textit{f}$^{1}$, 4\textit{f}$^3$\,5g$^{1}$,
		4\textit{f}$^3$\,6\textit{s}$^{1}$, 4\textit{f}$^3$\,6\textit{p}$^{1}$, 4\textit{f}$^3$\,6\textit{d}$^{1}$, 4\textit{f}$^3$\,6\textit{f}$^{1}$, 4\textit{f}$^3$\,6\textit{g}$^{1}$, 4\textit{f}$^3$\,7\textit{s}$^{1}$, 4\textit{f}$^3$\,7\textit{p}$^{1}$, 4\textit{f}$^3$\,7\textit{d}$^{1}$, 4\textit{f}$^3$\,7\textit{f}$^{1}$, 4\textit{f}$^3$\,7\textit{g}$^{1}$, 4\textit{f}$^3$\,8\textit{s}$^{1}$, \\4\textit{f}$^3$\,8\textit{p}$^{1}$, 4\textit{f}$^3$\,8\textit{d}$^{1}$, 4\textit{f}$^3$\,8\textit{f}$^{1}$, 4\textit{f}$^3$\,8\textit{g}$^{1}$, 4\textit{f}$^2$\,5\textit{d}$^2$, 4\textit{f}$^2$\,5\textit{d}$^{1}$\,6\textit{s}$^{1}$, 4\textit{f}$^2$\,5\textit{d}$^{1}$\,6\textit{p}$^{1}$, 4\textit{f}$^2$\,5\textit{d}$^{1}$\,6\textit{d}$^{1}$, 4\textit{f}$^2$\,6\textit{s}$^2$, 4\textit{f}$^2$\,6\textit{s}$^{1}$\,6\textit{p}$^{1}$, 4\textit{f}$^2$\,6\textit{s}$^{1}$\,6\textit{d}$^{1}$ \end{tabular}   \\\midrule
		         
		\multirow{4}{*}{$\ion{U}{ii}$} &
		-- $n=6$ & \begin{tabular}{l} 5\textit{f}$^3$\,7\textit{s}$^2$, 5\textit{f}$^5$, 5\textit{f}$^4$\,5g, 5\textit{f}$^4$\,6\textit{d}$^{1}$, 5\textit{f}$^4$\,6\textit{f}$^{1}$, 5\textit{f}$^4$\,6\textit{g}$^{1}$, 5\textit{f}$^3$6\textit{d}$^2$\end{tabular}   \\ 
		    & -- $n=7$ & \begin{tabular}{l} 5\textit{f}$^4$\,7\textit{s}$^{1}$, 5\textit{f}$^4$\,7\textit{p}$^{1}$, 5\textit{f}$^4$\,7\textit{d}$^{1}$, 5\textit{f}$^4$\,7\textit{f}$^{1}$, 5\textit{f}$^4$\,7\textit{g}$^{1}$, 5\textit{f}$^3$\,6\textit{d}$^{1}$\,7\textit{s}$^{1}$, 5\textit{f}$^3$\,6\textit{d}$^{1}$\,7\textit{p}$^{1}$, 5\textit{f}$^3$\,6\textit{d}$^{1}$\,7\textit{d}$^{1}$, 5\textit{f}$^3$\,7\textit{s}$^{1}$\,7\textit{p}$^{1}$, 5\textit{f}$^3$\,7\textit{s}$^{1}$\,7\textit{d}$^{1}$ \end{tabular} \\
		    & -- $n=8$ & \begin{tabular}{l} 5\textit{f}$^4$\,8\textit{s}$^{1}$, 5\textit{f}$^4$\,8\textit{p}$^{1}$, 5\textit{f}$^4$\,8\textit{d}$^{1}$, 5\textit{f}$^4$\,8\textit{f}$^{1}$, 5\textit{f}$^4$\,8\textit{g}$^{1}$ \end{tabular} \\
		    & -- $n=9$ & \begin{tabular}{l} 5\textit{f}$^4$\,9\textit{s}$^{1}$, 5\textit{f}$^4$\,9\textit{p}$^{1}$, 5\textit{f}$^4$\,9\textit{d}$^{1}$, 5\textit{f}$^4$\,9\textit{f}$^{1}$, 5\textit{f}$^4$\,9\textit{g}$^{1}$ \end{tabular} \\
		\midrule
		            
		\multirow{4}{*}{$\ion{U}{iii}$} &
		-- $n=6$ & \begin{tabular}{l} 5\textit{f}$^4$, 5\textit{f}$^3$\,5g$^{1}$, 5\textit{f}$^3$\,6\textit{d}$^{1}$, 5\textit{f}$^3$\,6\textit{f}$^{1}$, 5\textit{f}$^3$\,6\textit{g}$^{1}$, 5\textit{f}$^2$\,6\textit{d}$^2$\end{tabular}   \\ 
		    & -- $n=7$ & \begin{tabular}{l} 5\textit{f}$^3$\,7\textit{s}$^{1}$, 5\textit{f}$^3$\,7\textit{p}$^{1}$, 5\textit{f}$^3$\,7\textit{d}$^{1}$, 5\textit{f}$^3$\,7\textit{f}$^{1}$, 5\textit{f}$^3$\,7\textit{g}$^{1}$, 5\textit{f}$^2$\,6\textit{d}$^{1}$\,7\textit{s}$^{1}$, 5\textit{f}$^2$\,6\textit{d}$^{1}$\,7\textit{p}$^{1}$, 5\textit{f}$^2$\,6\textit{d}$^{1}$\,7\textit{d}$^{1}$, 5\textit{f}$^2$\,7\textit{s}$^2$, 5\textit{f}$^2$\,7\textit{s}$^{1}$\,7\textit{p}$^{1}$, 5\textit{f}$^2$\,7\textit{s}$^{1}$\,7\textit{d}$^{1}$ \end{tabular} \\
		    & -- $n=8$ & \begin{tabular}{l} 5\textit{f}$^3$\,8\textit{s}$^{1}$, 5\textit{f}$^3$\,8\textit{p}$^{1}$, 5\textit{f}$^3$\,8\textit{d}$^{1}$, 5\textit{f}$^3$\,8\textit{f}$^{1}$, 5\textit{f}$^3$\,8\textit{g}$^{1}$ \end{tabular} \\
		    & -- $n=9$ & \begin{tabular}{l} 5\textit{f}$^3$\,9\textit{s}$^{1}$, 5\textit{f}$^3$\,9\textit{p}$^{1}$, 5\textit{f}$^3$\,9\textit{d}$^{1}$, 5\textit{f}$^3$\,9\textit{f}$^{1}$, 5\textit{f}$^3$\,9\textit{g}$^{1}$ \end{tabular} \\
		\bottomrule
	\end{tabular}
\end{table*}

\section{Results}

\subsection{Nd atomic data}
\label{sec:Nd_atomic_data}

\begin{table*}
	\centering
	\caption{\ion{Nd}{ii} lowest level energies (in cm$^{-1}$) for $J$-$P$ split groups and comparison with those from NIST \citep{NIST_ASD}. Shown are the lowest bound states for each 2$J$-$P$ group, as well as the relative difference (in percent, columns $\Delta\%$) compared to NIST for the \FAC\, \ion{Nd}{ii} calculations involving 8 (both with and without potential optimisation), 15, 22 or 22~+~extra~CI configurations (see Table~\ref{tab:FAC_configs}).}
	\label{tab:ndii_comparison_nist}
    \begin{tabular}{ccrrrrrrrrrrr}
        \hline   
        2J    & P &  NIST &  \multicolumn{2}{|r|}{\FAC\, 8 conf. no opt.} & \multicolumn{2}{|r|}{\FAC\, 8 configurations} &  \multicolumn{2}{|r|}{\FAC\, 15 configurations} &  \multicolumn{2}{|r|}{\FAC\, 22 configurations} &  \multicolumn{2}{|r|}{\FAC\, 22 conf. + extra CI} \\
             &   &  lowest E &  lowest E  &    $\Delta$\% &  lowest E &    $\Delta$\% &  lowest E &   $\Delta$\% &  lowest E &    $\Delta$\% &  lowest E &    $\Delta$\%\\
        
        \hline
         7 &   +  &      0.00 & 10628.63 &     -- &     0.00 &  0.00 &        0.00 &  0.00   &       0.00 &  0.00 &       0.00 &  0.00 \\
         9 &   +  &    513.34 & 11437.85 & 2128.12 &   601.85 & 17.24 &      583.04 & 13.58   &     583.03 & 13.58 &     582.60 & 13.49 \\
        11 &   +  &   1470.14 & 12688.95 &  763.11 &  1431.41 &  2.63 &     1416.16 &  3.67   &    1416.16 &  3.67 &    1415.78 &  3.70 \\
        13 &   +  &   2585.52 & 14116.23 &  445.97 &  2422.04 &  6.32 &     2414.06 &  6.63   &    2414.07 &  6.63 &    2413.80 &  6.64 \\
        15 &   +  &   3802.02 & 15642.70 &  311.43 &  3536.74 &  6.98 &     3536.87 &  6.97   &    3536.89 &  6.97 &    3536.74 &  6.98 \\
        17 &   +  &   5085.76 & 17226.03 &  238.71 &  4747.88 &  6.64 &     4756.05 &  6.48   &    4756.08 &  6.48 &    4756.06 &  6.48 \\
        13 &   -- &   8009.99 &     0.00 &  100.00 &  4919.12 & 38.59 &     5195.95 & 35.13   &    3650.09 & 54.43 &    3645.25 & 54.49 \\
         3 &   +  &   8716.65 & 21778.79 &  149.85 & 11689.88 & 34.11 &    11676.38 & 33.95   &   11647.38 & 33.62 &   11565.20 & 32.68 \\
         5 &   +  &   8796.57 & 22195.96 &  152.33 & 12072.80 & 37.24 &    12039.59 & 36.87   &   12008.65 & 36.52 &   11925.96 & 35.58 \\
        19 &   +  &   9166.42 & 21649.58 &  136.18 &  9723.88 &  6.08 &     9986.61 &  8.95   &    9984.26 &  8.92 &    9914.43 &  8.16 \\
        15 &   -- &   9448.40 &  1391.09 &   85.28 &  6456.89 & 31.66 &     6733.58 & 28.73   &    5146.44 & 45.53 &    5143.85 & 45.56 \\
        11 &   -- &  10054.43 &  5845.92 &   41.86 & 10248.08 &  1.93 &    10503.46 &  4.47   &    9231.13 &  8.19 &    9214.29 &  8.36 \\
         9 &   -- &  10091.59 &  6022.28 &   40.32 & 10429.39 &  3.35 &    10692.66 &  5.96   &    9512.01 &  5.74 &    9471.40 &  6.15 \\
         1 &   +  &  10256.28 & 22771.73 &  122.03 & 13628.99 & 32.88 &    13634.60 & 32.94   &   13634.21 & 32.94 &   13633.86 & 32.93 \\
        21 &   +  &  10517.03 & 23384.47 &  122.35 & 10882.33 &  3.47 &    11148.54 &  6.00   &   11146.54 &  5.99 &   11076.00 &  5.31 \\
        17 &   -- &  10980.79 &  2942.02 &   73.21 &  8126.39 & 25.99 &     8402.95 & 23.48   &    6777.25 & 38.28 &    6775.91 & 38.29 \\
         7 &   -- &  12232.97 &  7162.72 &   41.45 & 11919.22 &  2.56 &    12187.71 &  0.37   &   10942.41 & 10.55 &   10928.36 & 10.66 \\
        19 &   -- &  12601.10 &  4604.02 &   63.46 &  9915.98 & 21.31 &    10192.43 & 19.11   &    8530.13 & 32.31 &    8529.41 & 32.31 \\
         5 &   -- &  13804.53 & 10225.02 &   25.93 & 14735.11 &  6.74 &    14992.54 &  8.61   &   14062.28 &  1.87 &   14013.44 &  1.51 \\
        21 &   -- &  14299.62 &  6303.31 &   55.92 & 11809.77 & 17.41 &    12086.11 & 15.48   &   10388.90 & 27.35 &   10388.43 & 27.35 \\
         3 &   -- &  15420.51 & 13264.48 &   13.98 & 18582.85 & 20.51 &    18826.29 & 22.09   &   17758.05 & 15.16 &   17739.32 & 15.04 \\
        23 &   -- &  16064.46 &  8003.61 &   50.18 & 13790.32 & 14.16 &    14066.59 & 12.44   &   12336.24 & 23.21 &   12335.78 & 23.21 \\
         1 &   -- &  33520.99 & 13140.21 &   60.80 & 18499.97 & 44.81 &    18741.91 & 44.09   &   17665.18 & 47.30 &   17647.56 & 47.35 \\
        \hline
    \end{tabular}
\end{table*}

\begin{table*}
	\centering
	\caption{\ion{Nd}{iii} lowest level energies (in cm$^{-1}$) for $J$-$P$ split groups and comparison with those from NIST \citep{NIST_ASD}. Shown are the lowest bound states for each 2$J$-$P$ group, as well as the relative difference (in percent, columns $\Delta\%$) compared to NIST for the \FAC\, \ion{Nd}{iii} calculations involving 8 (both with and without potential optimisation), 15, 22 or 22~+~extra~CI configurations (see Table~\ref{tab:FAC_configs}).}
	\label{tab:ndiii_comparison_nist}
    \begin{tabular}{ccrrrrrrrrrrr}
        \hline   
        2J    & P &  NIST &  \multicolumn{2}{|r|}{\FAC\, 8 conf. no opt.} &  \multicolumn{2}{|r|}{\FAC\, 8 configurations} &  \multicolumn{2}{|r|}{\FAC\, 15 configurations} &  \multicolumn{2}{|r|}{\FAC\, 22 configurations} &  \multicolumn{2}{|r|}{\FAC\, 22 conf. + extra CI} \\
             &   &  lowest E &  lowest E  &    $\Delta$\% &  lowest E &    $\Delta$\% &  lowest E &   $\Delta$\% &  lowest E &    $\Delta$\% &  lowest E &    $\Delta$\% \\
        \hline
         8 &   + &         0.00 &      0.00 &  0.00 &     0.00 &  0.00 &        0.00 &  0.00 &        0.00 &  0.00 &        0.00 &  0.00 \\
        10 &   + &      1137.83 &    793.68 & 30.25 &  1029.13 &  9.55 &     1026.62 &  9.77 &     1020.18 & 10.34 &     1021.76 & 10.20 \\
        12 &   + &      2387.62 &   1981.96 & 16.99 &  2189.48 &  8.30 &     2183.27 &  8.56 &     2169.79 &  9.12 &     2172.83 &  9.00 \\
        14 &   + &      3714.99 &   3585.11 &  3.50 &  3453.62 &  7.04 &     3442.57 &  7.33 &     3421.78 &  7.89 &     3426.03 &  7.78 \\
        16 &   + &      5093.43 &   5532.54 &  8.62 &  4797.84 &  5.80 &     4780.96 &  6.13 &     4752.83 &  6.69 &     4758.02 &  6.59 \\
        10 &  -- &     15262.54 &   4835.81 & 68.32 & 20809.88 & 36.35 &    20851.22 & 36.62 &    20783.77 & 36.18 &    16549.67 &  8.43 \\
        12 &  -- &     16938.52 &   4463.70 & 73.65 & 20105.72 & 18.70 &    20133.30 & 18.86 &    20067.46 & 18.47 &    15977.05 &  5.68 \\
        14 &  -- &     18656.76 &   6228.95 & 66.61 & 22152.75 & 18.74 &    22174.68 & 18.86 &    22106.86 & 18.49 &    17918.58 &  3.96 \\
         8 &  -- &     18884.13 &  10438.07 & 44.73 & 27807.08 & 47.25 &    27998.81 & 48.27 &    27967.67 & 48.10 &    23065.62 & 22.14 \\
         6 &  -- &     19211.44 &  11223.32 & 41.58 & 27578.69 & 43.55 &    27697.05 & 44.17 &    27630.07 & 43.82 &    23224.51 & 20.89 \\
        16 &  -- &     20411.38 &   8163.89 & 60.00 & 24342.01 & 19.26 &    24358.43 & 19.34 &    24288.61 & 19.00 &    20001.06 &  2.01 \\
        18 &  -- &     22197.53 &  10253.16 & 53.81 & 26640.54 & 20.02 &    26651.70 & 20.07 &    26579.88 & 19.74 &    22195.09 &  0.01 \\
        \hline
    \end{tabular}
\end{table*}

The lowest energy levels for each parity and $J$ for the largest calculations performed with \FAC\, and \HFR\, for \ion{Nd}{ii} and \ion{Nd}{iii} are shown in Tables \ref{tab:ndii_comparison_nist}, \ref{tab:ndiii_comparison_nist}, \ref{tab:ndii_comparison_nist_biggest}, and \ref{tab:ndiii_comparison_nist_biggest}. The relative difference from the data available in the NIST ASD \citep{NIST_ASD} is also evaluated as $\Delta_{X} = |E_{\mathrm{NIST}}-E_{X}|/E_{\mathrm{NIST}}$ with $X=\texttt{FAC}, \texttt{HFR}$. The \HFR\,calculations give rise to 13\,004\,380 and 6\,481\,846 lines computed for \ion{Nd}{ii} and \ion{Nd}{iii}, respectively, whereas the \FAC\,calculations yield a total number of 15\,031\,028 lines for \ion{Nd}{ii} and  1\,833\,759 for \ion{Nd}{iii}. Each level is labeled by the largest configuration in both $LS$ coupling, which are extracted from the NIST database, when available, and $jj$ coupling schemes, the latter obtained directly from the \texttt{FAC} output. While $LS$ labels use the usual $^{2S+1}L$ term symbols to represent the coupling of electrons, the label provided by \texttt{FAC} uses non-standard notation: a given subshell $i$ is denoted as ${(n_il_{i+}^{N_i})}_{2J_{i}}$ or ${(n_il_{i-}^{N_i})}_{2J_{i}}$. Here $n_i$ and $l_i$ represent the usual principal and angular momentum quantum numbers, $N_i$ represents the number of equivalent electrons within the same subshell while $2J_i$ denotes 2 times the total angular momentum that the electrons in the subshell couple to. Finally, we adopt the $+$ and $-$ notation to denote $j=l+1/2$ and $j=l-1/2$, respectively. The total angular momentum of the level given by the coupling of all subshells ${\mathbf J} = {\mathbf J}_1 + {\mathbf J}_2 + \dots$ is also indicated as a subscript.  As an example, the ground level of \ion{Nd}{iii} is labeled as $((4f_-^3)_{9} \, (4f_+^1)_{7})_8$ - in this case 3 equivalent electrons in a $4f_{5/2}$ subshell couple to give $2J_1=9$ while there is only one electron with $j=7/2$, and, therefore, $2J_2=7$. The level has a total angular momentum of $2J=8$.

\begin{table*}
    \caption{Excitation energies (in cm$^{-1}$) for the first 20 energy levels of \ion{Nd}{ii} calculated for the largest models computed with the \FAC\, and \HFR\, codes, with matching experimental values available in the NIST database \citep{NIST_ASD} . Relative differences with NIST data are also shown (columns  $\Delta_{\mathrm{FAC}} \%$ and $\Delta_{\mathrm{HFR}} \%$, in percent).}
    \label{tab:ndii_comparison_nist_biggest}
    \begin{tabular}{rlllrrrrr}
        \toprule
        $2J$ & $P$ & LS label & \FAC\, label &  $E_{\mathrm{NIST}}$ &  $E_{\FAC}$ &  $E_{\HFR}$ &  $\Delta_{\FAC} \%$ &  $\Delta_{\HFR} \%$ \\
        \midrule
        7 & + & $4f^{4}\,(^5\text{I})\,6s\,^6\text{I}$ & $(((4f_-^{3})_{9}\,(4f_+^{1})_{7})_{8}\,(6s_+^{1})_{1})_{7}$ & 0.00 & 0.00 & 5501.95 & -- & -- \\
        9 & + & $4f^{4}\,(^5\text{I})\,6s\,^6\text{I}$ & $(((4f_-^{3})_{9}\,(4f_+^{1})_{7})_{8}\,(6s_+^{1})_{1})_{9}$ & 513.33 & 582.59 & 3151.46 & 13.49 & 513.93 \\
        11 & + & $4f^{4}\,(^5\text{I})\,6s\,^6\text{I}$ & $(((4f_-^{3})_{9}\,(4f_+^{1})_{7})_{10}\,(6s_+^{1})_{1})_{11}$ & 1470.11 & 1415.75 & 3003.03 & 3.70 & 104.27 \\
        9 & + & $4f^{4}\,(^5\text{I})\,6s\,^4\text{I}$ & $(((4f_-^{3})_{9}\,(4f_+^{1})_{7})_{10}\,(6s_+^{1})_{1})_{9}$ & 1650.20 & 2010.85 & 4215.74 & 21.85 & 155.47 \\
        13 & + & $4f^{4}\,(^5\text{I})\,6s\,^6\text{I}$ & $(((4f_-^{2})_{8}\,(4f_+^{2})_{12})_{12}\,(6s_+^{1})_{1})_{13}$ & 2585.46 & 2413.74 & 0.00 & 6.64 & 100.00 \\
        11 & + & $4f^{4}\,(^5\text{I})\,6s\,^4\text{I}$ & $(((4f_-^{2})_{8}\,(4f_+^{2})_{12})_{12}\,(6s_+^{1})_{1})_{11}$ & 3066.76 & 3319.29 & 3722.90 & 8.23 & 21.40 \\
        15 & + & $4f^{4}\,(^5\text{I})\,6s\,^6\text{I}$ & $(((4f_-^{1})_{5}\,(4f_+^{3})_{15})_{14}\,(6s_+^{1})_{1})_{15}$ & 3801.93 & 3536.66 & 1521.20 & 6.98 & 59.99 \\
        11 & + & $4f^{4}\,(^5\text{I})\,5d\,^6\text{L}$ & $(((4f_-^{3})_{9}\,(4f_+^{1})_{7})_{8}\,(5d_-^{1})_{3})_{11}$ & 4437.56 & 6072.56 & 4094.33 & 36.84 & 7.73 \\
        13 & + & $4f^{4}\,(^5\text{I})\,6s\,^4\text{I}$ & $(((4f_-^{1})_{5}\,(4f_+^{3})_{15})_{14}\,(6s_+^{1})_{1})_{13}$ & 4512.49 & 4698.94 & 3809.67 & 4.13 & 15.58 \\
        17 & + & $4f^{4}\,(^5\text{I})\,6s\,^6\text{I}$ & $(((4f_-^{1})_{5}\,(4f_+^{3})_{15})_{16}\,(6s_+^{1})_{1})_{17}$ & 5085.64 & 4755.95 & 3146.14 & 6.48 & 38.14 \\
        13 & + & $4f^{4}\,(^5\text{I})\,5d\,^6\text{L}$ & $(((4f_-^{3})_{9}\,(4f_+^{1})_{7})_{10}\,(5d_-^{1})_{3})_{13}$ & 5487.65 & 6890.90 & 5021.58 & 25.57 & 8.49 \\
        15 & + & $4f^{4}\,(^5\text{I})\,6s\,^4\text{I}$ & $(((4f_-^{1})_{5}\,(4f_+^{3})_{15})_{16}\,(6s_+^{1})_{1})_{15}$ & 5985.58 & 6140.96 & 5372.38 & 2.60 & 10.24 \\
        9 & + & $4f^{4}\,(^5\text{I})\,5d\,^6\text{K}$ & $(((4f_-^{3})_{9}\,(4f_+^{1})_{7})_{8}\,(5d_-^{1})_{3})_{9}$ & 6005.27 & 7938.19 & 6459.96 & 32.19 & 7.57 \\
        15 & + & $4f^{4}\,(^5\text{I})\,5d\,^6\text{L}$ & $(((4f_-^{2})_{8}\,(4f_+^{2})_{12})_{12}\,(5d_-^{1})_{3})_{15}$ & 6637.43 & 7810.87 & 6500.77 & 17.68 & 2.06 \\
        11 & + & $4f^{4}\,(^5\text{I})\,5d\,^6\text{K}$ & $(((4f_-^{3})_{9}\,(4f_+^{1})_{7})_{10}\,(5d_-^{1})_{3})_{11}$ & 6931.80 & 8673.65 & 5393.85 & 25.13 & 22.19 \\
        7 & + & $4f^{4}\,(^5\text{I})\,5d\,^6\text{I}$ & $(((4f_-^{3})_{9}\,(4f_+^{1})_{7})_{8}\,(5d_-^{1})_{3})_{7}$ & 7524.73 & 10609.34 & 6799.30 & 40.99 & 9.64 \\
        17 & + & $4f^{4}\,(^5\text{I})\,5d\,^6\text{L}$ & $(((4f_-^{2})_{8}\,(4f_+^{2})_{12})_{12}\,(5d_+^{1})_{5})_{17}$ & 7868.91 & 8822.27 & 7126.82 & 12.12 & 9.43 \\
        13 & + & $4f^{4}\,(^5\text{I})\,5d\,^6\text{K}$ & $(((4f_-^{2})_{8}\,(4f_+^{2})_{12})_{12}\,(5d_-^{1})_{3})_{13}$ & 7950.07 & 9512.74 & 5176.15 & 19.66 & 34.89 \\
        13 & -- & $4f^{3}\,(^4\text{I}^{\mathrm{o}})\,5d^{2}\,(^3F)\,^6M^\mathrm{o}$ & $((4f_-^{3})_{9}\,(5d_-^{2})_{4})_{13}$ & 8009.81 & 3645.16 & 6719.31 & 54.49 & 16.11 \\
        9 & + & $4f^{4}\,(^5\text{I})\,5d\,^6\text{I}$ & $(((4f_-^{3})_{9}\,(4f_+^{1})_{7})_{10}\,(5d_-^{1})_{3})_{9}$ & 8420.32 & 11263.59 & 7373.54 & 33.77 & 12.43 \\
        \bottomrule
    \end{tabular}
\end{table*}

\begin{table*}
    \caption{Excitation energies (in cm$^{-1}$) for the first 20 energy levels of \ion{Nd}{iii} calculated for the largest models computed with the \FAC\, and \HFR\, codes, with matching experimental values available in the NIST database \citep{NIST_ASD} . Relative differences with NIST data are also shown (columns  $\Delta_{\mathrm{FAC}} \%$ and $\Delta_{\mathrm{HFR}} \%$, in percent).}
    \label{tab:ndiii_comparison_nist_biggest}
    \begin{tabular}{rlllrrrrr}
        \toprule
        $2J$ & $P$ & LS label & \FAC\, label &  $E_{\mathrm{NIST}}$ &  $E_{\FAC}$ &  $E_{\HFR}$ &  $\Delta_{\FAC} \%$ &  $\Delta_{\HFR} \%$ \\
        \midrule
        8 & + & $4f^{4}\,^5\text{I}$ & $((4f_-^{3})_{9}\,(4f_+^{1})_{7})_{8}$ & 0.0 & 0.00 & 0.00 & -- & -- \\
        10 & + & $4f^{4}\,^5\text{I}$ & $((4f_-^{3})_{9}\,(4f_+^{1})_{7})_{10}$ & 1137.8 & 1021.74 & 1241.96 & 10.20 & 9.15 \\
        12 & + & $4f^{4}\,^5\text{I}$ & $((4f_-^{2})_{8}\,(4f_+^{2})_{12})_{12}$ & 2387.6 & 2172.78 & 2606.19 & 9.00 & 9.16 \\
        14 & + & $4f^{4}\,^5\text{I}$ & $((4f_-^{1})_{5}\,(4f_+^{3})_{15})_{14}$ & 3714.9 & 3425.95 & 4051.72 & 7.78 & 9.07 \\
        16 & + & $4f^{4}\,^5\text{I}$ & $((4f_-^{1})_{5}\,(4f_+^{3})_{15})_{16}$ & 5093.3 & 4757.91 & 5548.06 & 6.58 & 8.93 \\
        10 & -- & $4f^{3}\,(^4\text{I}^{\mathrm{o}})\,5d\,^5\text{K}^{\mathrm{o}}$ & $((4f_-^{3})_{9}\,(5d_-^{1})_{3})_{10}$ & 15262.2 & 16549.28 & 4593.41 & 8.43 & 69.90 \\
        12 & -- & $4f^{3}\,(^4\text{I}^{\mathrm{o}})\,5d\,^5\text{K}^{\mathrm{o}}$ & $((4f_-^{3})_{9}\,(5d_-^{1})_{3})_{12}$ & 16938.1 & 15976.68 & 4352.91 & 5.68 & 74.30 \\
        14 & -- & $4f^{3}\,(^4\text{I}^{\mathrm{o}})\,5d\,^5\text{K}^{\mathrm{o}}$ & $(((4f_-^{2})_{8}\,(4f_+^{1})_{7})_{11}\,(5d_-^{1})_{3})_{14}$ & 18656.3 & 17918.16 & 6256.33 & 3.96 & 66.47 \\
        8 & -- & $4f^{3}\,(^4\text{I}^{\mathrm{o}})\,5d\,^5\text{I}^{\mathrm{o}}$ & $((4f_-^{3})_{9}\,(5d_-^{1})_{3})_{8}$ & 18883.7 & 23065.09 & 9187.74 & 22.14 & 51.35 \\
        6 & -- & $4f^{3}\,(^4\text{I}^{\mathrm{o}})\,5d\,^5\text{H}^{\mathrm{o}}$ & $((4f_-^{3})_{9}\,(5d_-^{1})_{3})_{6}$ & 19211.0 & 23223.97 & 9972.34 & 20.89 & 48.09 \\
        8 & --& $4f^{3}\,(^4\text{I}^{\mathrm{o}})\,5d\,^5\text{H}^{\mathrm{o}}$ & $((4f_-^{3})_{9}\,(5d_+^{1})_{5})_{8}$ & 20144.3 & 23889.54 & 10779.05 & 18.59 & 46.49 \\
        10 & -- & $4f^{3}\,(^4\text{I}^{\mathrm{o}})\,5d\,^5\text{I}^{\mathrm{o}}$ & $((4f_-^{3})_{9}\,(5d_+^{1})_{5})_{10}$ & 20388.9 & 22656.18 & 9828.27 & 11.12 & 51.80 \\
        16 & -- & $4f^{3}\,(^4\text{I}^{\mathrm{o}})\,5d\,^5\text{K}^{\mathrm{o}}$ & $(((4f_-^{2})_{8}\,(4f_+^{1})_{7})_{11}\,(5d_+^{1})_{5})_{16}$ & 20410.9 & 20000.60 & 8284.38 & 2.01 & 59.41 \\
        10 & -- & $4f^{3}\,(^4\text{I}^{\mathrm{o}})\,5d\,^5\text{H}^{\mathrm{o}}$ & $(((4f_-^{2})_{8}\,(4f_+^{1})_{7})_{11}\,(5d_-^{1})_{3})_{10}$ & 21886.8 & 24612.33 & 10761.02 & 12.45 & 50.83 \\
        12 & -- & $4f^{3}\,(^4\text{I}^{\mathrm{o}})\,5d\,^5\text{I}^{\mathrm{o}}$ & $(((4f_-^{2})_{8}\,(4f_+^{1})_{7})_{11}\,(5d_-^{1})_{3})_{12}$ & 22047.8 & 18298.85 & 6333.83 & 17.00 & 71.27 \\
        18 & -- & $4f^{3}\,(^4\text{I}^{\mathrm{o}})\,5d\,^5\text{K}^{\mathrm{o}}$ & $(((4f_-^{1})_{5}\,(4f_+^{2})_{12})_{13}\,(5d_+^{1})_{5})_{18}$ & 22197.0 & 22194.58 & 10410.99 & 0.01 & 53.10 \\
        14 & -- & $4f^{3}\,(^4\text{I}^{\mathrm{o}})\,5d\,^5\text{I}^{\mathrm{o}}$ & $(((4f_-^{2})_{8}\,(4f_+^{1})_{7})_{11}\,(5d_+^{1})_{5})_{14}$ & 22702.9 & 20147.62 & 8143.06 & 11.26 & 64.13 \\
        12 & -- & $4f^{3}\,(^4\text{I}^{\mathrm{o}})\,5d\,^5\text{H}^{\mathrm{o}}$ & $(((4f_-^{2})_{8}\,(4f_+^{1})_{7})_{11}\,(5d_-^{1})_{3})_{12}$ & 23819.3 & 23973.50 & 11011.64 & 0.65 & 53.77 \\
        14 & -- & $4f^{3}\,(^4\text{I}^{\mathrm{o}})\,5d\,$ & $(((4f_-^{1})_{5}\,(4f_+^{2})_{12})_{13}\,(5d_-^{1})_{3})_{14}$ & 24003.2 & 26522.01 & 13134.57 & 10.49 & 45.28 \\
        16 & -- & $4f^{3}\,(^4\text{I}^{\mathrm{o}})\,5d\,^5\text{I}^{\mathrm{o}}$ & $(((4f_-^{1})_{5}\,(4f_+^{2})_{12})_{13}\,(5d_+^{1})_{5})_{16}$ & 24686.4 & 22083.05 & 10010.19 & 10.55 & 59.45 \\
        \bottomrule
    \end{tabular}
\end{table*}

Associated with the density of levels, and as reported in previous works \citep[][]{2019ApJS..240...29G,2020ApJS..248...17R, 2022MNRAS.517..281G} strong mixing between configurations is anticipated for both lanthanides and actinides. In such cases, the assignment of configurations to each individual level becomes more complicated. For this reason, levels present in the NIST ASD \citep{NIST_ASD} database were identified and matched to the theoretical values primarily based on their $J$, parity, and their position within a $J$-$P$ group. Although comparisons with the available experimental atomic energy levels should be reliable for the first few levels, the large gaps in the experimental data prevent direct comparisons of more excited levels, as direct matching to the calculated levels becomes unreliable for the reasons mentioned above. Using this approach, a total of 611 levels were identified for the calibrated 22 config + extra CI calculation of \ion{Nd}{ii} in the NIST ASD. The average relative difference to experimental data is of $\overline{\Delta_{\texttt{FAC}}\%}=9.21\%$ and $\overline{\Delta_{\texttt{HFR}}\%}=18.35\%$ with mean energy differences of $\overline{\Delta_{\texttt{FAC}}}=1970 \text{\,cm}^{-1}$ and $\overline{\Delta_{\texttt{HFR}}}=3838 \text{\,cm}^{-1}$.

For \ion{Nd}{iii}, the 29 experimental levels available in the NIST ASD represent only 0.38\% of the total number of theoretical levels (15230) computed with both codes used in this work. Therefore, it is difficult to precisely assess the accuracy of these calculations, in particular for the higher energy levels. We observe, however, a larger disparity between the two sets of calculations, with a mean accuracy of the \FAC\, calculations similar to the \ion{Nd}{ii} case, ($\overline{\Delta_{\texttt{FAC}}\%}=9.65\%$, corresponding to $\overline{\Delta_{\texttt{FAC}}}=1975 \text{\,cm}^{-1}$), while a larger deviation from the experimental data is found for the \HFR\ energies ($\overline{\Delta_{\texttt{HFR}}\%}=49.41\%$ and $\overline{\Delta_{\texttt{HFR}}}=9234 \text{\,cm}^{-1}$). Following the level identification described above, a one-to-one matching between \texttt{FAC} and \texttt{HFR} was possible for 27330 levels of \ion{Nd}{ii} and 9668 levels of \ion{Nd}{iii}. An average relative difference (evaluated as $|E_{\mathrm{FAC}}-E_\mathrm{HFR}|/E_{\mathrm{FAC}}$) of 10.65\% and of 22.54\% was found for the singly and doubly charged states, respectively.

\begin{figure*}
    \includegraphics[width=\textwidth]{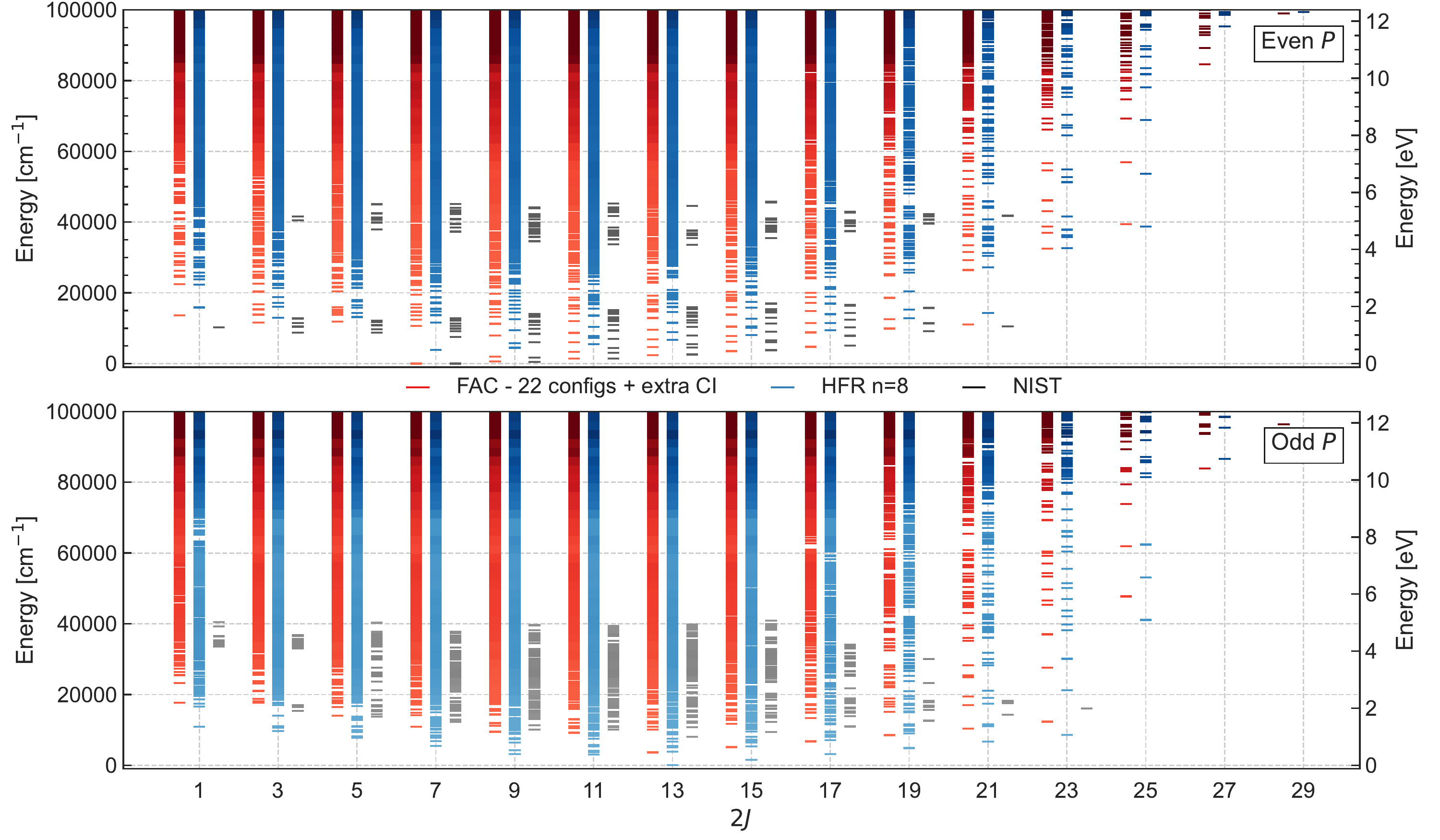}
    \caption{Energy levels for \ion{Nd}{ii} for even (top) and odd (bottom) parity for the models with the larger number of configurations using both \FAC\, and the \HFR\, codes, depicted in red and blue horizontal lines, respectively. For comparison, data compiled in the NIST database \citep{NIST_ASD} is also shown in black. Darker colors show a higher density of levels in that region.}  
    \label{fig:NdII_levels_jpi}
\end{figure*}

\begin{figure*}
    \includegraphics[width=\textwidth]{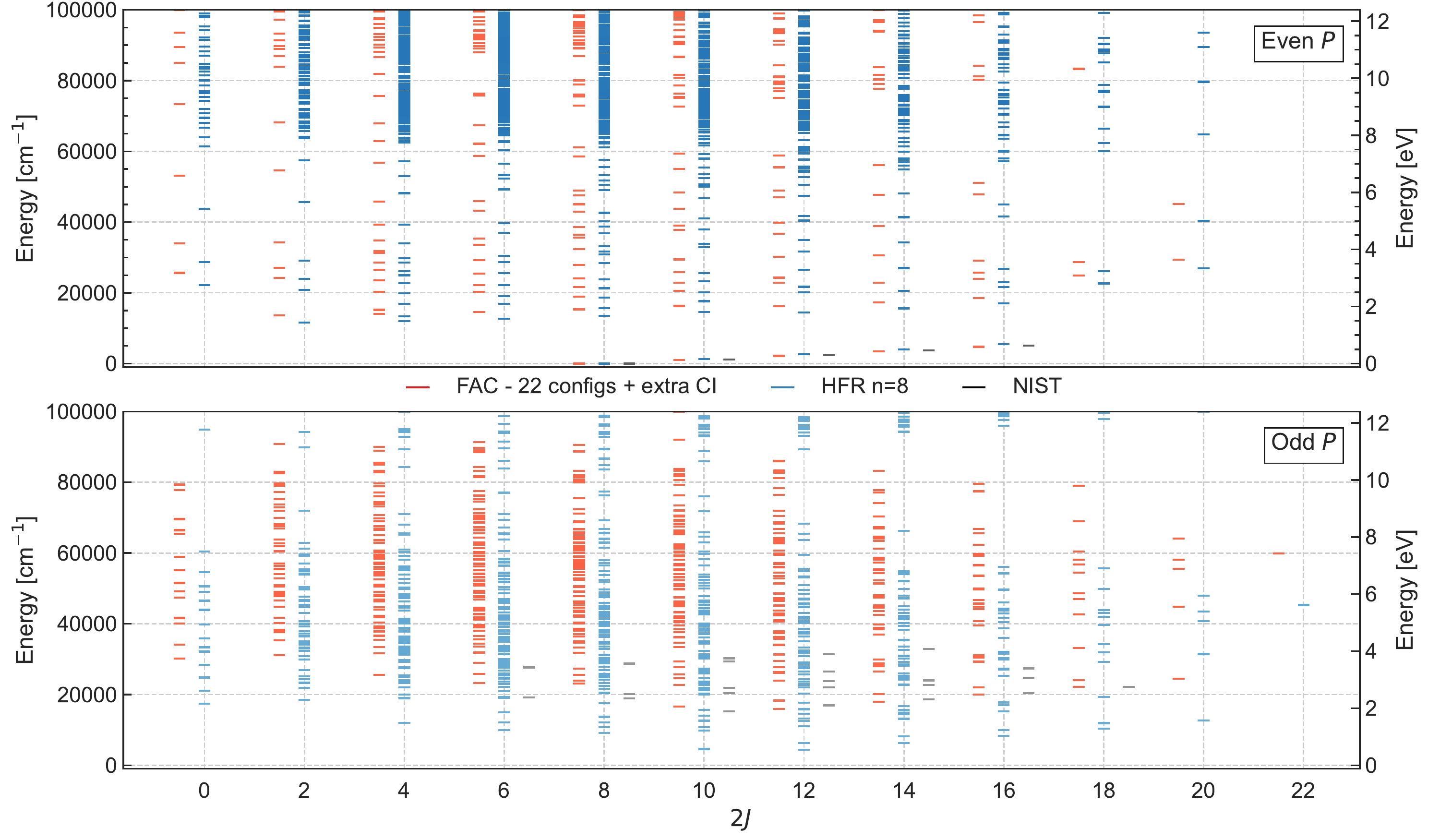}
    \caption{Energy levels for \ion{Nd}{iii} for even (top) and odd (bottom) parity for the models which include the larger number of configurations using both \FAC\, and the \HFR\, codes, depicted in red and blue horizontal lines, respectively. For comparison, data compiled in the NIST database \citep{NIST_ASD} is also shown in black. Darker colors show a higher density of levels in that region.}
    \label{fig:NdIII_levels_jpi}
\end{figure*}

A visual representation of the energy levels for the largest calculations achieved with \FAC\, and \HFR\, as well as their comparison to the data available in the NIST ASD can be found in Figure~\ref{fig:NdII_levels_jpi} for \ion{Nd}{ii} and Figure~\ref{fig:NdIII_levels_jpi} for \ion{Nd}{iii}.  In the case of the singly ionised ion, while some differences are found between the two codes used, particularly for $2J = 7$ and $2J=23$, the dispersion of levels is similar, especially for the low-lying levels, where the level density is the lowest. Larger differences between \HFR\ and \FAC\ results are observed for \ion{Nd}{iii}, particularly for the even parity. \HFR\,calculations indeed predict a higher level density than \FAC\, in the 60~000 -- 10~0000~cm$^{-1}$ range: $20.23 \ \text{levels}/1000\ \text{cm}^{-1}$ compared to a level density of $7.30 \ \text{levels}/1000\ \text{cm}^{-1} $ from \FAC. This may be explained by the presence of the three extra configurations considered in the \HFR\ calculation with respect to the \FAC\ one (\textit{i.e.}, $4f^2\,6s^2$, $4f^2\,5d^1\,6d^1$ and $4f^2\,6s^1\,6d^1$), which give rise to 1200 extra levels that lie between 70~000 and 200~000 cm$^{-1}$. For the odd parity states the disparity is not as large. However, our \HFR\,calculations give rise to levels with energies lower than those of \FAC, with a constant shift of about $\sim 13~000$~cm$^{-1}$. This could be explained by differences in the optimization of the orbitals in each of the codes. While in \FAC\ the optimization of the potential is based on a restricted number of configurations chosen among all the configurations included in the physical model, the average energies are minimised for the whole set of \HFR\ configurations. This difference between the two methods is thought to explain the lower energies computed by \HFR\ for the levels belonging to configurations that are not optimised in the \FAC\ calculations. The resulting effect on the opacity cannot be ignored (see Section~\ref{sec:opacities} for a further discussion).

To assess the reliability of the relevant atomic data supplied, we have studied the convergence of the energy levels with the multiple models calculated using the \FAC\, code. The lowest level energies for every $J$ and parity symmetries for the different calculations performed were compared to the data available on the NIST ASD. Results for singly- and doubly-ionised Nd are shown in Tables~\ref{tab:ndii_comparison_nist} and \ref{tab:ndiii_comparison_nist}, respectively. In both tables we see that the potential optimization of \texttt{FAC} (described in Section \ref{sec:fac}) has the biggest impact on the energies levels, with improvements of $211 \%$ for \ion{Nd}{ii} and of $19.5 \%$ for \ion{Nd}{iii}.

The impact of the inclusion of a higher number of configurations is particularly noticeable for \ion{Nd}{iii}, where the mean deviation from NIST recommended values of the lowest state of each $J$-$P$ group decreased from $19.54 \%$ to $8.06 \%$, with a greater impact on the odd parity states (where there was an overall improvement of about $20\%$).

Larger deviations (of 40--50\% in some cases) were found for the lowest energy levels of each $J$-$P$ group of \ion{Nd}{ii}, even for the largest CI calculations based on 30~configurations (\texttt{22 configurations + extra CI}). The mean deviation to NIST data also increased from $16.20 \%$ to $19.30 \%$. We should highlight, however, that the impact of the level energy precision on the opacities is not the same for all levels. Lower excited levels, in particular, should have a larger effect under the assumption of local thermodynamic equilibrium (LTE).

\begin{figure*}
    \includegraphics[width=\textwidth]{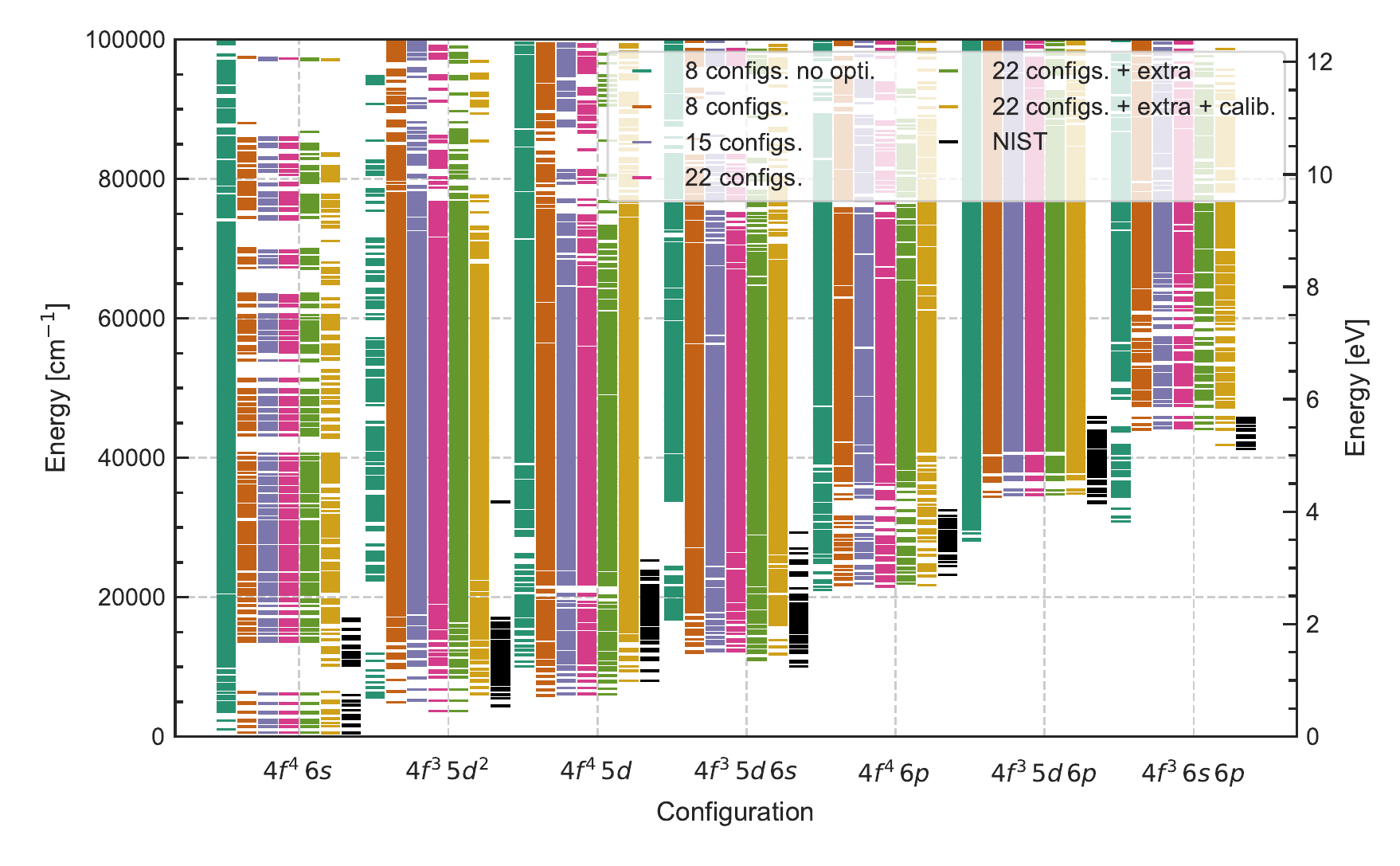}
    \caption{Energy levels of each configuration for \ion{Nd}{ii}. Black horizontal lines show the data from the NIST ASD \citep{NIST_ASD}. Coloured horizontal lines show the calculated data from our 8, 15, 22 and 30 configurations calculations as well as the calibrated data for 30 configurations. The calibrations were performed within each P-J group.}
    \label{fig:NdII_levels}
\end{figure*}

While direct effects on the opacity will be characterised in \ref{sec:opacities}, a more in-depth convergence analysis of the calculated energy levels using \FAC\, was carried out for the case of \ion{Nd}{ii} where the average accuracy (with respect to NIST ASD experimental data) of the lowest energy levels is computed for different calculations. Figure \ref{fig:NdII_levels} shows the effect of the different models in the resulting energy levels of the 7 lowest configurations of \ion{Nd}{ii}. In particular, we highlight the large effect of the optimization of the local central potential, described in Section \ref{sec:fac}. This optimization has, by far, the biggest impact on the atomic data, and illustrates how sensitive the \FAC\, calculations are to the chosen local potential and to the choice of the FMC. In any case, it is important to ensure the convergence of the energy levels with the inclusion of higher degrees of correlation through the inclusion of more configurations in the CSF basis set used in the CI scheme. 

\begin{figure}
    \centering
    \includegraphics[width=\columnwidth]{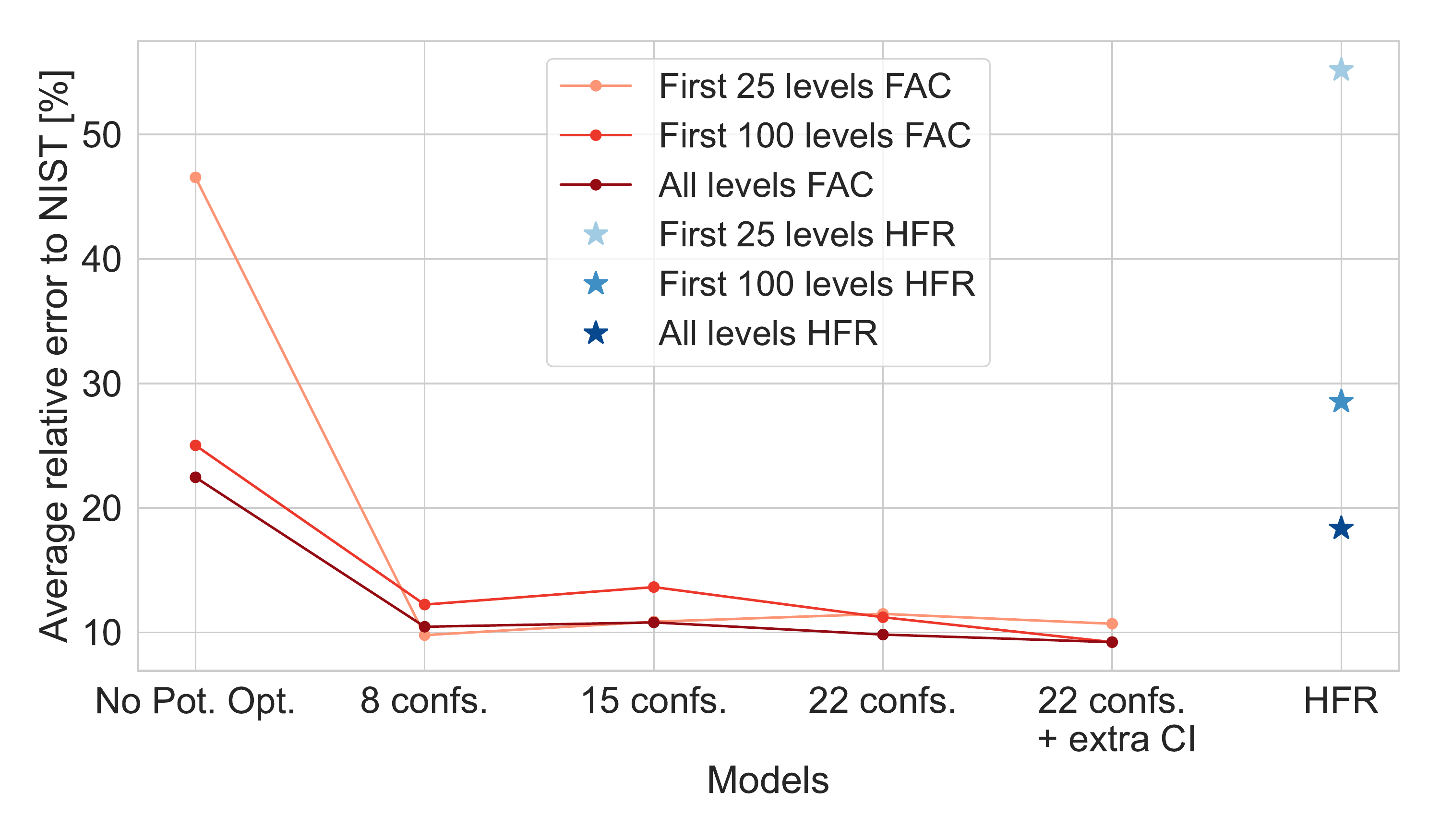}
    \caption{Average relative error to the data available in the NIST ASD \citep{NIST_ASD} for sets of levels (25, 50, and All) for the different models computed with \FAC\, and for the \HFR\, calculation for \ion{Nd}{ii}. ``No Pot. Opt.'' corresponds to an 8 configuration model that does not use the \texttt{FAC} potential optimization, contrarily to all other models.}
    \label{fig:convergence_levels_Ndii}
\end{figure}

Nonetheless, the impact of the additional configurations on the energy levels is on average very minor (typically only a few per cent), as can be seen from Figure \ref{fig:convergence_levels_Ndii}. While the effect on the accuracy seems to be irregular, indicating that only 30 configurations may still not be sufficient to achieve convergence, we note that the inclusion of the extra configurations in the largest model, for which transitions were not computed, still resulted in a noticeable effect mainly on the more excited levels. The effect of the calibration applied in the last model has a more noticeable effect for the first even ($4\textit{f}^4\, 6s^1$) and odd ($4\textit{f}^4\, 5d^1$) configurations since the constant shift applied is adjusted to match the energy of the lowest levels of each $J$-$P$ block. 

When compared to the \FAC\ calculations, larger deviations from experimental values are found for the AELs of \ion{Nd}{ii} computed with \HFR\,(greater than $50 \%$) This is partly due to the fact that the predicted \HFR\ ground state differs from observation and from the \FAC\ calculations, whatever the correlation model (see Table~\ref{tab:ndii_comparison_nist_biggest}). It is also worth mentioning that configuration average energy adjustments were tested to match the \ion{Nd}{ii} \HFR\ predicted ground level to the observation. While it was possible to obtain the observed ground state in the \HFR\ atomic data when proceeding this way, the impact on the opacities computed with this set of data was negligible \citep{2023EPJD.in.prep}.

\subsection{U atomic data}
\label{sec:U_atomic_data}
Just as for the Nd atomic data, the lowest energy levels for each $J$ and parity for the most complex models used with both \FAC\, and \HFR\, for \ion{U}{ii} and \ion{U}{iii} are respectively shown in Tables \ref{tab:uii_comparison_nist}, \ref{tab:uiii_comparison_nist},  \ref{tab:uii_comparison_scasa_biggest} and \ref{tab:uiii_comparison_scasa_biggest}. As no data, besides for the ground state, is available in the NIST ASD, the calculation obtained for the considered uranium ions are compared to data available in the Selected Constants Energy Levels and Atomic Spectra of Actinides (SCASA). The same methodology used for NIST data is used here to evaluate the relative difference from the data available in the SCASA database with $\Delta_{\text{FAC}} = |E_{\text{SCASA}}-E_{\text{FAC}}|/E_{\text{SCASA}}$ and $\Delta_{\text{HFR}} = |E_{\text{SCASA}}-E_{\text{HFR}}|/E_{\text{SCASA}}$. With the \HFR\,code, 12\,758\,946 and 7\,177\,574 lines are obtained for \ion{U}{ii} and \ion{U}{iii}, respectively, whereas the \FAC\,calculations give rise to 11\,605\,793 lines for \ion{U}{ii} and 2\,549\,511 lines for \ion{U}{iii}.

As explained in Section~\ref{sec:Nd_atomic_data}, energy levels from the SCASA database were identified and matched to our computed values based on their $J$-value, parity $P$, and their position within a $J$-$P$ group. $LS$-labels are, in this case, taken directly from the SCASA database. Level identification for \ion{U}{ii} and \ion{U}{iii} has been carried out from the experimental work of Palmer and Engleman Jr. \citep{palmer1984}, together with calculations of Blaise and co-workers \citep{Blaise1984, Blaise1987, Blaise1992}.  Unfortunately, a number of $J$-values are still ambiguous, with a few levels from the original analysis still to be confirmed, particularly for the doubly-charged ion. Theoretical interpretation of these levels is still lacking for several of the measured levels. For this reason, unidentified levels or levels which predicted leading component has a percentage of less than $25 \%$ are only labelled by their electronic configuration, i.e. omitting the (unidentified) term symbol.

\texttt{FAC} labelling follows the same format as the one introduced in Section~\ref{sec:Nd_atomic_data}. Some discrepancies in the configuration of the leading component of the eigenvector have been observed between labels. While some variation is to be expected due to the use of different coupling schemes, we notice the large uncertainties in the identification of levels for the uranium spectra, which are partly related to the strong mixing and high level density,  even at lower energies. Such issues may not directly affect the calculation of opacities, but they should be considered, particularly when assessing the quality and reliability of atomic data.

As for Nd, except for the first few levels, the lack of data for more excited levels makes the matching of the latter to our calculated values less reliable. For \ion{U}{ii}, the average relative difference to SCASA data is of $\overline{\Delta_{\texttt{FAC}}\%}=28.44\%$ and $\overline{\Delta_{\texttt{HFR}}\%}=31.34 \%$ with mean energy differences of $\overline{\Delta_{\texttt{FAC}}}=7988.46 \text{\,cm}^{-1}$  and $\overline{\Delta_{\texttt{HFR}}}=8867.60 \text{\,cm}^{-1}$. In the case of \ion{U}{iii}, agreement between the reliably matched SCASA data seems to favour the \HFR\,calculation, for which no calibration on experimental data was performed and for which we find an average relative difference of $\overline{\Delta_{\texttt{HFR}}\%}=25.10 \%$, and mean energy different to the data of $\overline{\Delta_{\texttt{HFR}}}=6041.28 \text{\,cm}^{-1}$ compared to the results obtained with \FAC, $\overline{\Delta_{\texttt{FAC}}\%}=31.51\%$ and $\overline{\Delta_{\texttt{FAC}}}=8275.58 \text{\,cm}^{-1}$. We find average relative differences (measured as $|E_{\text{FAC}}-E_{\text{HFR}}|/E_{\text{FAC}}$) of $12.36\%$ for \ion{U}{ii}, with 26 844 identified and matched levels between the calculations of the two codes, and of $15.64 \%$ for \ion{U}{iii}, with 9427 matched levels.

Comparison of the lowest levels in each $J$-$P$ group with the levels from the SCASA database is also shown in Tables~\ref{tab:uii_comparison_nist} and \ref{tab:uiii_comparison_nist} for the multiple models used in the \FAC\, calculations. The biggest relative differences are found for the lowest states of \ion{U}{ii}, due to the fact our \FAC\,results predict a different ground level to the one reported in the SCASA database. Similar to Nd, the inclusion of extra configurations has a bigger impact on the energy levels for the doubly-ionised case, with agreement to SCASA data improving to $7.46 \%$ for \ion{U}{ii} and $183.93 \%$ for \ion{U}{iii} when going from 8 to 30 configurations in the CSF basis sets, all \FAC\,CI calculations including the optimization of the central potential. Just looking at the lower $J$-$P$ states of each group, we do observe an improvement of $33 \%$ for \ion{U}{ii} despite the fact that the \FAC\ ground state differs from SCASA. For \ion{U}{iii}, the optimization process was necessary to fix the wrong ground state. On average, while the agreement with measured data worsens with the optimization for the doubly-ionised ion (much due to the odd $2J=12$ state), the energies tend to converge to values closer to the experimental ones with the increasing number of configurations added in the model.

\begin{figure*}
    \includegraphics[width=\textwidth]{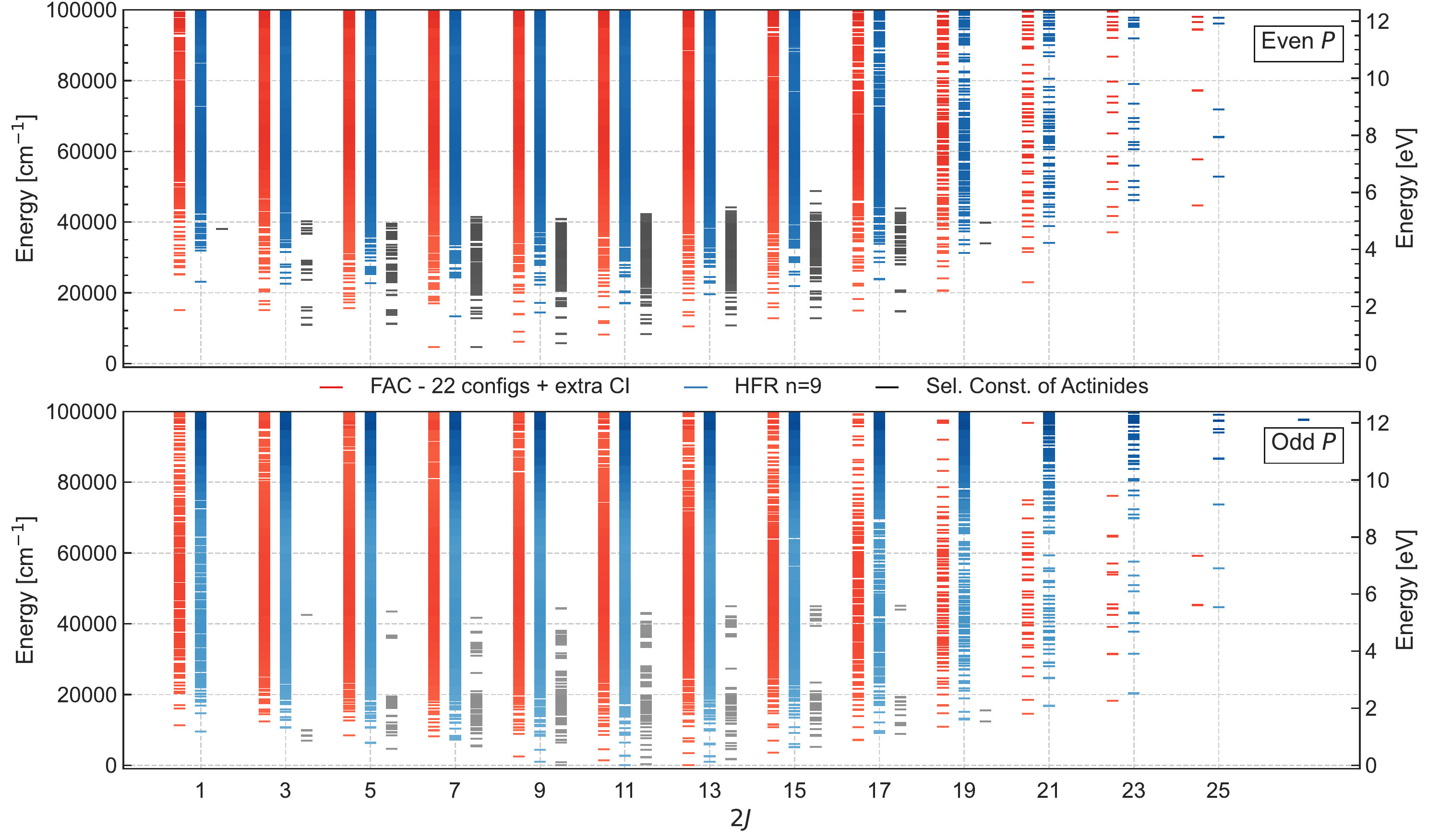}
    \caption{Energy levels for \ion{U}{ii} for even (top) and odd (bottom) parity for the models with the larger number of configurations using both \FAC\, and the \HFR\, codes, depicted in red and blue horizontal lines, respectively. Black horizontal lines shown experimental data from the Selected Constants Energy Levels and Atomic Spectra of Actinides \citep{Blaise1992}. Darker colors show a higher density of levels in that region.}  
    \label{fig:UII_levels_jpi}
\end{figure*}

\begin{figure*}
    \includegraphics[width=\textwidth]{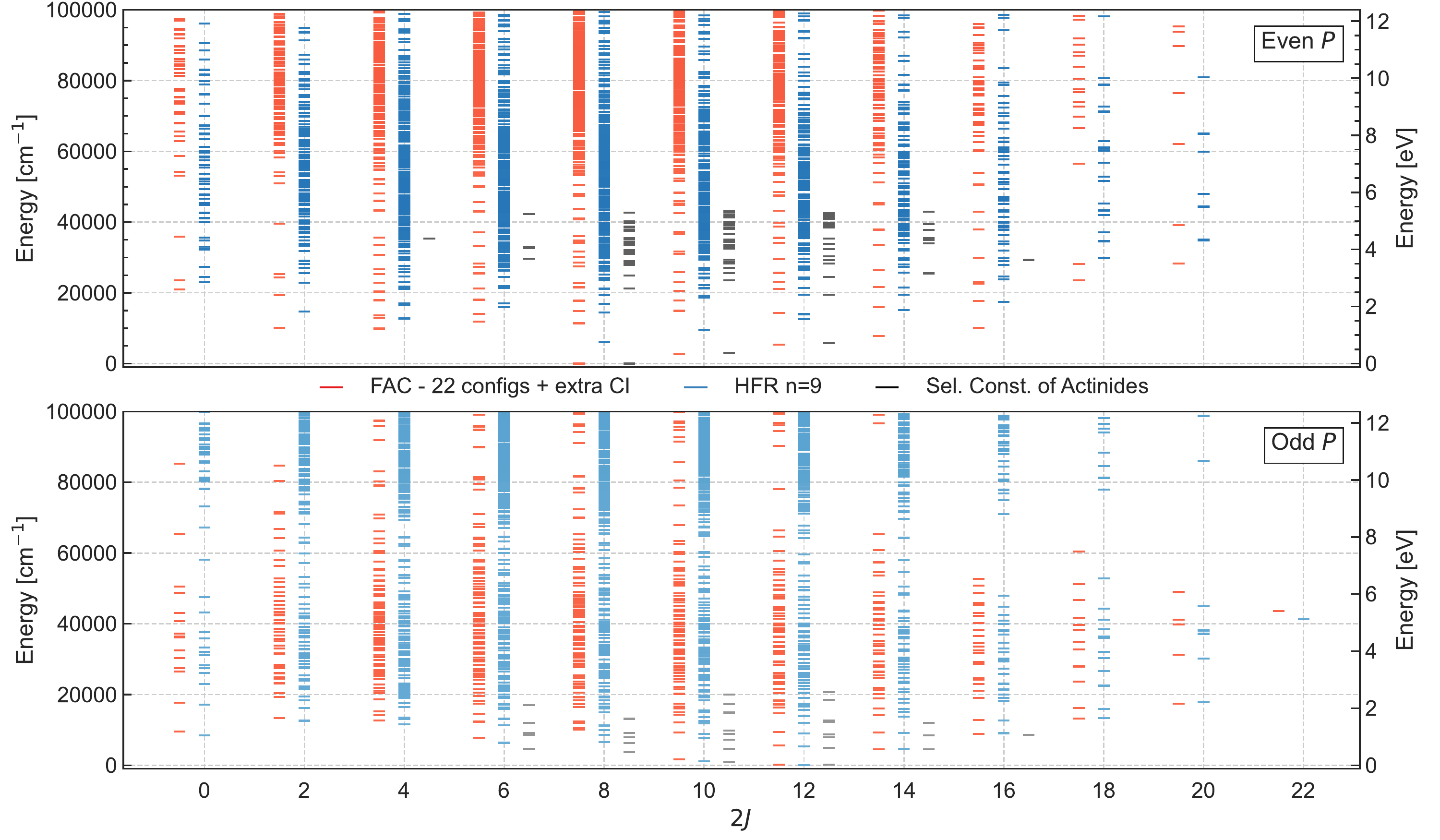}
    \caption{Energy levels for \ion{U}{iii} for even (top) and odd (bottom) parity for the models with the larger number of configurations using both \FAC\, and the \HFR\, codes, depicted in red and blue horizontal lines, respectively. Black horizontal lines shown experimental data from the Selected Constants Energy Levels and Atomic Spectra of Actinides \citep{Blaise1992}. Darker colours show a higher density of levels in that region.}  
    \label{fig:UIII_levels_jpi}
\end{figure*}

The levels obtained with the biggest computations performed with both \FAC\, and \HFR\, are visually represented in Figures~\ref{fig:UII_levels_jpi} and \ref{fig:UIII_levels_jpi} for \ion{U}{ii} and \ion{U}{iii}, respectively. A comparison with the available data from SCASA is also shown, highlighting the lack of available data that we intend to fill in the present work. The dispersion of levels obtained in both cases are in good agreement for singly-ionised uranium, even if a few differences can be observed, especially for the highest values of $J$ considered. Nevertheless, in the case of doubly-ionised uranium, the level density predicted by \HFR\, seems to be higher than the \FAC\, one below 50000 cm$^{-1}$ for even parity, while we find a much better agreement for the odd parity states. As for \ion{Nd}{iii}, the higher level density predicted by \HFR\ in the even parity can partially be explained by the three extra configurations included in the \HFR\ calculation and not in the \FAC\ model (\textit{i.e.}, $5f^2\,7s^2$, $5f^2\,6d^1\,7d^1$ and $5f^2\,7s^1\,7d^1$), which lead to 1200 additional levels of energies comprised between 4 000 and 130 000 cm$^{-1}$. In addition, as discussed in Section \ref{sec:Nd_atomic_data}, the fact that the average energies of all configurations are optimised in \HFR\ (while in \FAC\ the optimization of the potential is based on selected configurations of the model) is consistent with the higher level density predicted by \HFR\ at lower energies. The differences between the codes and their effects on the opacity will be further discussed in Section \ref{sec:opacities}. 

Despite the differences between the two sets of results obtained with \FAC\ and \HFR\ respectively, it is important to note that both calculations confirm the previous results from \citet{2022Atoms..10...18S} where a higher level density of low-lying levels of \ion{U}{iii} was found when compared to \ion{Nd}{iii}. Even with an identical atomic structure, this effect can be understood by the higher diffuseness of the $5f$ shell when compared to $4f$ \citep{1981tass.book.....C}. The larger radii (e.g. $5f$ compared to $4f$) increase the $5f -  \{6d, \, 7s, \, 7p\}$ overlap, relatively to the overlap of $4f$ with the outer shells, increasing the levels density. While this effect is not strong enough to produce noticeable effects for \ion{U}{ii}, similar effects are expected to appear in other actinide elements, particularly for ions with a lower number of electrons in the $5f$ shell.

As for \ion{Nd}{ii}, a configuration average energy adjustment procedure was tried for \ion{U}{ii} and \ion{U}{iii} to match the \HFR\ predicted ground state to the observed one, but the impact on the computed opacities was insignificant \citep{2023EPJD.in.prep}.

\begin{table*}
	\centering
	\caption{\ion{U}{ii} lowest level energies (in cm$^{-1}$) for $J$-$P$ split groups and comparison with those from Selected Constants Energy Levels and Atomic Spectra of Actinides (SCASA) \citep{Blaise1992}. Shown are the lowest bound states for each $J$-P group, as well as the relative difference (in percent, columns $\Delta\%$) compared to SCASA for the \FAC\, \ion{U}{ii} calculations involving 8 (both with and without potential optimisation), 15, 22 or 22~+~extra~CI configurations (see Table~\ref{tab:FAC_configs}).}
	\label{tab:uii_comparison_nist}
    \begin{tabular}{ccrrrrrrrrrrr}
        \hline   
        2J & P & SCASA & \multicolumn{2}{|r|}{\FAC\, 8 conf. no opt.} &  \multicolumn{2}{|r|}{\FAC\, 8 configurations} &  \multicolumn{2}{|r|}{\FAC\, 15 configurations} &  \multicolumn{2}{|r|}{\FAC\, 22 configurations} &  \multicolumn{2}{|r|}{\FAC\, 22 conf. + extra CI} \\
             &   &  lowest E &  lowest E  &    $\Delta$\% &  lowest E &    $\Delta$\% &  lowest E &   $\Delta$\% &  lowest E &    $\Delta$\% &  lowest E &    $\Delta$\% \\
        \hline
         9 &  -- &      0.00 &  1581.75 &    -- &  2690.67 &    --  &  2583.47 &    --  &  2530.90 &    --  &  2543.55 &    --  \\
        11 &  -- &    289.04 &     0.00 & 100.00 &  1539.06 & 432.47 &  1418.58 & 390.79 &  1361.33 & 370.98 &  1377.76 & 376.67 \\
        13 &  -- &   1749.12 &  1496.35 &  14.45 &     0.00 & 100.00 &     0.00 & 100.00 &     0.00 & 100.00 &     0.00 & 100.00 \\
         7 &   + &   4663.80 & 17818.42 & 282.06 &  5005.40 &   7.32 &  4907.60 &   5.23 &  4816.78 &   3.28 &  4650.32 &   0.29 \\
         5 &  -- &   4706.27 &  6547.31 &  39.12 &  8591.43 &  82.55 &  8489.08 &  80.38 &  8438.38 &  79.30 &  8448.19 &  79.51 \\
        15 &  -- &   5259.65 &  5984.35 &  13.78 &  3557.13 &  32.37 &  3563.44 &  32.25 &  3566.56 &  32.19 &  3565.85 &  32.20 \\
         7 &  -- &   5401.50 &  7983.44 &  47.80 &  8198.13 &  51.77 &  8189.17 &  51.61 &  8184.50 &  51.52 &  8185.10 &  51.53 \\
         9 &   + &   5716.45 & 19473.06 & 240.65 &  6488.03 &  13.50 &  6383.24 &  11.66 &  6285.96 &   9.96 &  6120.64 &   7.07 \\
         3 &  -- &   7017.17 & 11050.09 &  57.47 & 12504.18 &  78.19 & 12425.99 &  77.08 & 12388.86 &  76.55 & 12398.95 &  76.69 \\
        11 &   + &   8347.69 & 23092.05 & 176.63 &  8644.45 &   3.56 &  8526.57 &   2.14 &  8417.87 &   0.84 &  8259.73 &   1.05 \\
        17 &  -- &   8853.75 & 10405.92 &  17.53 &  7159.69 &  19.13 &  7168.31 &  19.04 &  7172.55 &  18.99 &  7171.52 &  19.00 \\
        13 &   + &  10740.26 & 22864.69 & 112.89 & 10934.15 &   1.81 & 10804.83 &   0.60 & 10686.29 &   0.50 & 10533.14 &   1.93 \\
         3 &   + &  10987.20 & 26381.81 & 140.11 & 15512.31 &  41.19 & 15441.00 &  40.54 & 15377.99 &  39.96 & 15151.98 &  37.91 \\
         5 &   + &  11252.34 & 26509.91 & 135.59 & 16046.61 &  42.61 & 15975.66 &  41.98 & 15913.13 &  41.42 & 15682.78 &  39.37 \\
        19 &  -- &  12350.36 & 14756.61 &  19.48 & 10857.56 &  12.09 & 10867.18 &  12.01 & 10871.88 &  11.97 & 10870.67 &  11.98 \\
        15 &   + &  12862.15 & 28490.21 & 121.50 & 13200.38 &   2.63 & 13063.08 &   1.56 & 12937.85 &   0.59 & 12788.09 &   0.58 \\
        17 &   + &  14796.72 & 31430.08 & 112.41 & 15397.98 &   4.06 & 15255.96 &   3.10 & 15127.04 &   2.23 & 14979.71 &   1.24 \\
        19 &   + &  33932.80 & 38421.68 &  13.23 & 20599.29 &  39.29 & 20608.87 &  39.27 & 20613.79 &  39.25 & 20623.11 &  39.22 \\
         1 &   + &  38053.38 & 27912.85 &  26.65 & 15465.59 &  59.36 & 15387.03 &  59.56 & 15317.40 &  59.75 & 15098.02 &  60.32 \\
        \hline
    \end{tabular}
\end{table*}

\begin{table*}
	\centering
	\caption{\ion{U}{iii} lowest level energies (in cm$^{-1}$) for $J$-$P$ split groups and comparison with those from Selected Constants Energy Levels and Atomic Spectra of Actinides (SCASA) \citep{Blaise1992}. Shown are the lowest bound states for each 2$J$-$P$ group, as well as the relative difference (in percent, columns $\Delta\%$) compared to SCASA for the \FAC\, \ion{U}{iii} calculations involving 8 (both with and without potential optimisation), 15, 22 or 22~+~extra~CI configurations (see Table~\ref{tab:FAC_configs}).}
	\label{tab:uiii_comparison_nist}
    \begin{tabular}{ccrrrrrrrrrrr}
        \hline   
        2J & P & SCASA & \multicolumn{2}{|r|}{\FAC\, 8 conf. no opt.} &  \multicolumn{2}{|r|}{\FAC\, 8 configurations} &  \multicolumn{2}{|r|}{\FAC\, 15 configurations} &  \multicolumn{2}{|r|}{\FAC\, 22 configurations} &  \multicolumn{2}{|r|}{\FAC\, 22 conf. + extra CI} \\
             &   &  lowest E &  lowest E  &    $\Delta$\% &  lowest E &    $\Delta$\% &  lowest E &   $\Delta$\% &  lowest E &    $\Delta$\% &  lowest E &    $\Delta$\% \\
        \hline
         8 &   + &     0.00 &  9974.20 &    --  &     0.00 &    0.00 &     0.00 &    0.00 &     0.00 &    0.00 &     0.00 &   0.00 \\
        12 &  -- &   210.26 &     0.00 & 100.00  &  3466.13 & 1548.46 &  3776.21 & 1695.93 &  3567.56 & 1596.70 &   140.29 &  33.28 \\
        10 &  -- &   885.33 &  1740.26 &  96.57  &  5283.70 &  496.80 &  5500.12 &  521.25 &  5335.41 &  502.65 &  1706.64 &  92.77 \\
        10 &   + &  3036.60 & 13565.59 & 346.74  &  2721.26 &   10.38 &  2685.83 &   11.55 &  2658.73 &   12.44 &  2686.43 &  11.53 \\
         8 &  -- &  3743.96 &  7495.65 & 100.21  & 13777.56 &  267.99 & 13881.63 &  270.77 & 13791.92 &  268.38 & 10121.67 & 170.35 \\
        14 &  -- &  4504.54 &  4770.13 &   5.90  &  8107.11 &   79.98 &  8425.70 &   87.05 &  8192.71 &   81.88 &  4501.51 &   0.07 \\
         6 &  -- &  4611.93 &  7629.25 &  65.42  & 11075.97 &  140.16 & 11208.56 &  143.03 & 11057.58 &  139.76 &  7733.36 &  67.68 \\
        12 &   + &  5719.42 & 11917.82 & 108.37  &  5377.10 &    5.99 &  5314.84 &    7.07 &  5267.35 &    7.90 &  5313.23 &   7.10 \\
        16 &  -- &  8649.88 &  9515.32 &  10.01  & 12732.63 &   47.20 & 13056.05 &   50.94 & 12802.24 &   48.00 &  8873.87 &   2.59 \\
        14 &   + & 25507.79 & 17166.55 &  32.70  &  7875.92 &   69.12 &  7797.72 &   69.43 &  7738.25 &   69.66 &  7796.31 &  69.44 \\
        16 &   + & 29310.58 & 22148.87 &  24.43  & 10214.43 &   65.15 & 10129.29 &   65.44 & 10064.90 &   65.66 & 10131.22 &  65.43 \\
         6 &   + & 29668.40 & 20836.10 &  29.77  & 12016.11 &   59.50 & 12056.34 &   59.36 & 12095.56 &   59.23 & 11926.25 &  59.80 \\
         4 &   + & 35309.11 & 18196.19 &  48.47  &  9962.94 &   71.78 & 10039.56 &   71.57 & 10108.43 &   71.37 &  9894.50 &  71.98 \\
        \hline
    \end{tabular}
\end{table*}

\begin{table*}
    \caption{Excitation energies (in cm$^{-1}$) for the first 20 energy levels of \ion{U}{ii} calculated for the largest models computed with the \FAC\, and \HFR\, codes, with matching experimental values available in the Selected Constants Energy Levels and Atomic Spectra of Actinides (SCASA) \citep{Blaise1992} . Relative differences with SCASA data are also shown ( columns  $\Delta_{\mathrm{FAC}} \%$ and  $\Delta_{\mathrm{HFR}} \%$, in percent).}
    \label{tab:uii_comparison_scasa_biggest}
    \begin{tabular}{rlllrrrrr}
        \toprule
         $2J$ & $P$ & LS label &                                                                             FAC label &  $E_{SCASA}$ &  $E_{FAC}$ &  $E_{HFR}$ &  $\Delta_{FAC} \%$ &  $\Delta_{HFR} \%$ \\
        \midrule
        9 & -- & $5f^{3}\, (^{4}\text{I}^{\mathrm{o}}) \, 7s^2 \, ^{4}\text{I}^{\mathrm{o}}$ & $(((5f_-^{3})_{9}\,(6d_-^{1})_{3})_{10}\,(7s_+^{1})_{1})_{9}$ & 0.00 & 2543.49 & 956.71 & -- & -- \\
        11 & -- & $5f^{3}\, (^{4}\text{I}^{\mathrm{o}}) \, 6d \, (^{2}d^{\mathrm{o}}) \, 7s \, ^{6}\text{L}^{\mathrm{o}}$ & $(((5f_-^{3})_{9}\,(6d_-^{1})_{3})_{12}\,(7s_+^{1})_{1})_{11}$ & 289.04 & 1377.73 & 0.00 & 376.65 & 100.00 \\
        9 & -- & $5f^{3}\, (^{4}\text{I}^{\mathrm{o}}) \, 6d \, (^{2}d^{\mathrm{o}}) \, 7s \, ^{6}\text{K}^{\mathrm{o}}$ & $(5f_-^{3})_{9} \, 7s^2$ & 914.77 & 8902.13 & 4444.29 & 873.16 & 385.84 \\
        13 & -- & $5f^{3}\, (^{4}\text{I}^{\mathrm{o}}) \, 6d \, (^{2}d^{\mathrm{o}}) \, 7s \, ^{6}\text{L}^{\mathrm{o}}$ & $((5f_-^{3})_{9}\,(6d_-^{2})_{4})_{13}$ & 1749.12 & 0.00 & 1035.54 & 100.00 & 40.80 \\
        11 & -- & $5f^{3}\, (^{4}\text{I}^{\mathrm{o}}) \, 6d \, (^{2}d^{\mathrm{o}}) \, 7s \, ^{6}\text{K}^{\mathrm{o}}$ & $(((5f_-^{3})_{9}\,(6d_-^{1})_{3})_{10}\,(7s_+^{1})_{1})_{11}$ & 2294.70 & 4548.04 & 2701.12 & 98.20 & 17.71 \\
        11 & -- &  $5f^{3}\, (^{4}\text{I}^{\mathrm{o}}) \, 7s^2 \, ^{4}\text{I}^{\mathrm{o}}$ & $((((5f_-^{2})_{8}\,(5f_+^{1})_{7})_{11}\,(6d_-^{1})_{3})_{12}\,(7s_+^{1})_{1})_{11}$ & 4420.87 & 8558.11 & 6449.03 & 93.58 & 45.88 \\
        13 & -- &  $5f^{3}\, (^{4}\text{I}^{\mathrm{o}}) \, 6d^2 \, (^{3}f^{\mathrm{o}}) \, ^{6}\text{M}^{\mathrm{o}}$ & $(((5f_-^{3})_{9}\,(6d_-^{1})_{3})_{12}\,(7s_+^{1})_{1})_{13}$ & 4585.43 & 3439.26 & 2569.10 & 25.00 & 43.97 \\
        7 & + & $5f^{4}\, (^{5}\text{I}) \, 7s \, ^{6}\text{I}$	 & $((5f_-^{4})_{8}\,(7s_+^{1})_{1})_{7}$ & 4663.80 & 4650.21 & 13381.88 & 0.29 & 186.93 \\
        5 & -- & $5f^{3}\, (^{4}\text{I}^{\mathrm{o}}) \, 6d \, (^{2}d^{\mathrm{o}}) \, 7s \, ^{6}\text{H}^{\mathrm{o}}$ & $(((5f_-^{3})_{9}\,(6d_-^{1})_{3})_{6}\,(7s_+^{1})_{1})_{5}$ & 4706.27 & 8448.00 & 6357.92 & 79.51 & 35.09 \\
        15 & -- & $5f^{3}\, (^{4}\text{I}^{\mathrm{o}}) \, 6d \, (^{2}d^{\mathrm{o}}) \, 7s \, ^{6}\text{L}^{\mathrm{o}}$ & $(((5f_-^{2})_{8}\,(5f_+^{1})_{7})_{11}\,(6d_-^{2})_{4})_{15}$ & 5259.65 & 3565.77 & 5132.22 & 32.21 & 2.42 \\
        7 & -- & $5f^{3}\, 6d \, 7s$ & $((5f_-^{3})_{9}\,(6d_-^{2})_{4})_{7}$ & 5401.50 & 8184.91 & 7300.43 & 51.53 & 35.16 \\
        13 & -- & $5f^{3}\, (^{4}\text{I}^{\mathrm{o}}) \, 6d \, (^{2}d^{\mathrm{o}}) \, 7s \, ^{6}\text{K}^{\mathrm{o}}$ & $((((5f_-^{2})_{8}\,(5f_+^{1})_{7})_{11}\,(6d_-^{1})_{3})_{14}\,(7s_+^{1})_{1})_{13}$ & 5526.75 & 6764.89 & 5938.35 & 22.40 & 7.45 \\
        7 & -- & $5f^{3}\, (^{4}\text{I}^{\mathrm{o}}) \, 6d \, (^{2}d^{\mathrm{o}}) \, 7s \, ^{6}\text{I}^{\mathrm{o}}$ & $(((5f_-^{3})_{9}\,(6d_-^{1})_{3})_{6}\,(7s_+^{1})_{1})_{7}$ & 5667.33 & 9889.73 & 7850.66 & 74.50 & 38.52 \\
        9 & + & $5f^{4}\, (^{5}\text{I}) \, 7s \, ^{6}\text{I}$ & $((5f_-^{4})_{8}\,(7s_+^{1})_{1})_{9}$ & 5716.45 & 6120.50 & 14441.32 & 7.07 & 152.63 \\
        11 & -- & $5f^{3}\, (^{4}\text{I}^{\mathrm{o}}) \, 6d \, (^{2}d^{\mathrm{o}}) \, 7s \, ^{6}\text{K}^{\mathrm{o}}$ & $((5f_-^{3})_{9}\,(6d_-^{2})_{4})_{11}$ & 5790.64 & 9733.74 & 8614.23 & 68.09 & 48.76 \\
        13 & -- & $5f^{3}\, (^{4}\text{I}^{\mathrm{o}}) \, 6d \, (^{2}d^{\mathrm{o}}) \, 7s \, ^{4}L^{\mathrm{o}}$ & $((((5f_-^{2})_{8}\,(5f_+^{1})_{7})_{11}\,(6d_-^{1})_{3})_{12}\,(7s_+^{1})_{1})_{13}$ & 6283.43 & 7617.94 & 6134.38 & 21.24 & 2.37 \\
        9 & -- & $5f^{3}\, (^{4}\text{I}^{\mathrm{o}}) \, 6d \, (^{2}d^{\mathrm{o}}) \, 7s \, ^{6}\text{I}^{\mathrm{o}}$ & $(5f_-^{3})_{9} \, 7s^2$ & 6445.04 & 9907.09 & 8310.37 & 53.72 & 28.94 \\
        3 & -- & $5f^{3}\, (^{4}f^{\mathrm{o}}) \, 7s^2 \, ^{4}f^{\mathrm{o}}$ & $(((5f_-^{3})_{9}\,(6d_+^{1})_{5})_{4}\,(7s_+^{1})_{1})_{3}$ & 7017.17 & 12398.66 & 10691.21 & 76.69 & 52.36 \\
        9 & -- & $5f^{3}\, 6d \, 7s$ & $((5f_-^{3})_{9}\,(6d_-^{2})_{4})_{9}$ & 7166.63 & 10701.76 & 8778.63 & 49.33 & 22.49 \\
        7 & -- & $5f^{3}\, (^{4}\text{I}^{\mathrm{o}}) \, 6d \, (^{2}d^{\mathrm{o}}) \, 7s \, ^{4}\text{H}^{\mathrm{o}}$ & $(((5f_-^{3})_{9}\,(6d_-^{1})_{3})_{8}\,(7s_+^{1})_{1})_{7}$ & 7547.37 & 10883.51 & 8365.00 & 44.20 & 10.83 \\
        \bottomrule
    \end{tabular}
\end{table*}

\begin{table*}
    \caption{Excitation energies (in cm$^{-1}$) for the first 20 energy levels of \ion{U}{iii} calculated for the largest models computed with the \FAC\, and \HFR\, codes, with matching experimental values available in the Selected Constants Energy Levels and Atomic Spectra of Actinides (SCASA) \citep{Blaise1992} . Relative differences with SCASA data are also shown ( columns  $\Delta_{\mathrm{FAC}} \%$ and  $\Delta_{\mathrm{HFR}} \%$, in percent).}
    \label{tab:uiii_comparison_scasa_biggest}
    \begin{tabular}{rlllrrrrr}
        \toprule
        $2J$ & $P$ & LS label & FAC label &  $E_{NIST}$ &  $E_{FAC}$ &  $E_{HFR}$ &  $\Delta_{FAC} \%$ &  $\Delta_{HFR} \%$ \\
        \midrule
        8 & + & $5f^4 \, ^{5}\text{I}$ & $(5f_-^{4})_{8}$ & 0.00 & 0.00 & 6079.87 & -- & -- \\
        12 & -- & $5f^3 \, (^{4}\text{I}^{\mathrm{o}}) \, 6d \, ^{5}\text{L}^{\mathrm{o}}$ & $((5f_-^{3})_{9}\,(6d_-^{1})_{3})_{12}$ & 210.26 & 140.28 & 0.00 & 33.28 & 100.00 \\
        10 & -- & $5f^3 \, (^{4}\text{I}^{\mathrm{o}}) \, 6d \, ^{5}\text{K}^{\mathrm{o}}$ & $((5f_-^{3})_{9}\,(6d_-^{1})_{3})_{10}$ & 885.33 & 1706.60 & 1140.77 & 92.76 & 28.85 \\
        10 & + & $5f^4 \, ^{5}\text{I}$ & $((5f_-^{3})_{9}\,(5f_+^{1})_{7})_{10}$ & 3036.60 & 2686.36 & 9592.05 & 11.53 & 215.88 \\
        8 & -- & $5f^3 \, (^{4}\text{I}^{\mathrm{o}}) \, 7s\, ^{5}\text{I}^{\mathrm{o}}$& $((5f_-^{3})_{9}\,(6d_+^{1})_{5})_{8}$ & 3743.96 & 10121.44 & 6545.84 & 170.34 & 74.84 \\
        14 & -- & $5f^3 \, (^{4}\text{I}^{\mathrm{o}}) \, 6d \, ^{5}\text{L}^{\mathrm{o}}$ & $(((5f_-^{2})_{8}\,(5f_+^{1})_{7})_{11}\,(6d_-^{1})_{3})_{14}$ & 4504.54 & 4501.41 & 4625.95 & 0.07 & 2.70 \\
        6 & -- & $5f^3 \, 6d \, ^{5}\text{H}^{\mathrm{o}}$& $((5f_-^{3})_{9}\,(6d_-^{1})_{3})_{6}$ & 4611.93 & 7733.19 & 6349.06 & 67.68 & 37.67 \\
        10 & -- & $5f^3 \, (^{4}\text{I}^{\mathrm{o}}) \, 7s \, ^{5}\text{I}^{\mathrm{o}}$ & $((5f_-^{3})_{9}\,(6d_+^{1})_{5})_{10}$ & 4717.55 & 9284.92 & 7720.17 & 96.82 & 63.65 \\
        12 & -- & $5f^3 \, (^{4}\text{I}^{\mathrm{o}}) \, 6d \, ^{5}\text{K}^{\mathrm{o}}$ & $(((5f_-^{2})_{8}\,(5f_+^{1})_{7})_{11}\,(6d_-^{1})_{3})_{12}$ & 4939.62 & 5635.02 & 5373.54 & 14.08 & 8.78 \\
        12 & + & $5f^4 \, ^{5}\text{I}$ & $((5f_-^{2})_{8}\,(5f_+^{2})_{12})_{12}$ & 5719.42 & 5313.11 & 12602.68 & 7.10 & 120.35 \\
        8 & -- & $5f^3 \, (^{4}\text{I}^{\mathrm{o}}) \, 6d \, ^{5}\text{I}^{\mathrm{o}}$ & $((5f_-^{3})_{9}\,(7s_+^{1})_{1})_{8}$ & 6286.39 & 10477.60 & 8613.84 & 66.67 & 37.02 \\
        10 & -- & $5f^3 \, 6d$ & $((5f_-^{3})_{9}\,(7s_+^{1})_{1})_{10}$ & 7288.21 & 12179.88 & 8870.90 & 67.12 & 21.72 \\
        12 & -- & $5f^3 \, 6d$ & $(((5f_-^{2})_{8}\,(5f_+^{1})_{7})_{9}\,(6d_-^{1})_{3})_{12}$ & 7894.46 & 9431.47 & 8952.61 & 19.47 & 13.40 \\
        8 & -- & $5f^3 \, 6d$ & $((5f_-^{3})_{9}\,(6d_-^{1})_{3})_{8}$ & 7894.69 & 11980.70 & 9963.73 & 51.76 & 26.21 \\
        14 & -- & $5f^3 \, (^{4}\text{I}^{\mathrm{o}}) \, 6d \, ^{5}\text{K}^{\mathrm{o}}$ & $(((5f_-^{2})_{8}\,(5f_+^{1})_{7})_{11}\,(6d_+^{1})_{5})_{14}$ & 8437.71 & 9274.11 & 9099.53 & 9.91 & 7.84 \\
        6 & -- & $5f^3 \, 6d$ & $((5f_-^{3})_{9}\,(6d_+^{1})_{5})_{6}$ & 8568.51 & 12328.23 & 11276.53 & 43.88 & 31.60 \\
        16 & -- & $5f^3 \, (^{4}\text{I}^{\mathrm{o}}) \, 6d \, ^{5}\text{L}^{\mathrm{o}}$ & $(((5f_-^{2})_{8}\,(5f_+^{1})_{7})_{11}\,(6d_+^{1})_{5})_{16}$ & 8649.88 & 8873.66 & 9075.73 & 2.59 & 4.92 \\
        12 & -- & $5f^3 \, (^{4}\text{I}^{\mathrm{o}}) \, 7s \, ^{5}\text{I}^{\mathrm{o}}$ & $(((5f_-^{2})_{8}\,(5f_+^{1})_{7})_{11}\,(6d_+^{1})_{5})_{12}$ & 8778.31 & 14691.30 & 11961.99 & 67.36 & 36.27 \\
        10 & -- & $5f^3 \, (^{4}\text{I}^{\mathrm{o}}) \, 7s \, ^{3}\text{I}^{\mathrm{o}}$ & $(((5f_-^{2})_{8}\,(5f_+^{1})_{7})_{11}\,(6d_+^{1})_{5})_{10}$ & 8816.33 & 14297.63 & 11912.01 & 62.17 & 35.11 \\
        8 & -- & $5f^3 \, 6d$ & $(((5f_-^{2})_{8}\,(5f_+^{1})_{7})_{11}\,(6d_-^{1})_{3})_{8}$ & 9113.22 & 12427.52 & 11246.88 & 36.37 & 23.41 \\
        \bottomrule
    \end{tabular}
\end{table*}

\subsection{Bound-Bound Opacities}
\label{sec:opacities}

For the ions considered in this study (\ion{Nd}{ii}, \ion{Nd}{iii}, \ion{U}{ii}, \ion{U}{iii}), we compute the bound-bound opacities for conditions expected for the ejecta of NS mergers $\sim$1\,day after coalescence. However, we caution the reader that the presented opacities serve as illustrations and as a means to compare with data provided by other groups rather than direct input for radiative transfer models. Line-by-line opacities, as well as frequency dependent and grey opacities, are sensitive to the composition of the ejecta, and thus, the ionisation balance for the given composition. Approximating a more realistic (e.g. solar) composition with just a single element (Nd or U) would yield opacities that could be different by several orders of magnitude.

There exist a number of frequency-dependent opacity formalisms in the literature. In this study, we focus on the widely used expansion opacity \citep[see ][]{1993ApJ...412..731E, 2013ApJ...774...25K, 2013ApJ...775..113T}
\begin{equation}
    \kappa_{\mathrm{exp}} (\lambda) = \frac{1}{ct\rho}\sum_l \frac{\lambda_l}{\Delta\lambda}\left(1-e^{-\tau_l}\right),
    \label{equ:expansion_opacity}
\end{equation}
where $t$ is the time since merger, $\rho$ is the ejecta density, $\Delta\lambda$ is the bin width and $\lambda_l$ is the transition wavelength with Sobolev optical depth
\begin{equation}
    \tau_l = \frac{\pi e^2}{m_e c } f_l n_l \lambda_l t .
    \label{equ:sobolev_optical_depth}
\end{equation}
For the Sobolev optical depth, $f_l$ is the oscillator strength of the line with transition wavelength $\lambda_l$ and the corresponding lower level number density $n_l$. The number densities of the lower levels can be computed by assuming LTE holds using the Saha ionisation
\begin{equation}
    \frac{n_i}{n_{i-1}} = \frac{Z_i(T)g_e}{Z_{i-1}(T)n_e}e^{-E_{\mathrm{ion}}/k_B T}
\end{equation}
and Boltzmann excitation equations

\begin{equation}
    n_k = \frac{g_k}{Z_i(T)}e^{-E_k/k_B T}n_i, 
    \label{equ:boltzmann_excitation}
\end{equation}
where the indices $i$ and $k$ indicate ionisation states and level numbers, respectively, $g_k$ is the multiplicity of the level $(2J_k+1)$, $Z_i(T)$ are the partition functions and $E_k$ is the level energy. For the bin width, we chose 100\,\AA, but emphasise that neither the absolute scale nor the general shape of the expansion opacity depends on the bin width. Choosing a larger bin width has the same effect as smoothing the curves. In the literature, a bin width of $10\,$\AA\,is often assumed. For infinitesimally small $\Delta\lambda\rightarrow 0$ the expansion opacity reduces to the Sobolev line opacity. The Sobolev approximation \citep{1960mes..book.....S} is well justified for NS merger ejecta as the velocity gradient (homologous expansion) is steep and the corresponding expansion velocities ($\approx 0.1c$) greatly exceed the thermal line width (of order 1\,km\,s$^{-1}$), assuming no particularly strong hydrodynamic instabilities are in effect. While for arbitrary velocity gradients, the Sobolev optical depth depends on $\frac{dr}{dv}$, for a homologously expanding atmosphere ($\frac{dr}{dv}=t$) it simplifies to Equation~\ref{equ:sobolev_optical_depth}.

\begin{figure*}
    \includegraphics[width=\textwidth]{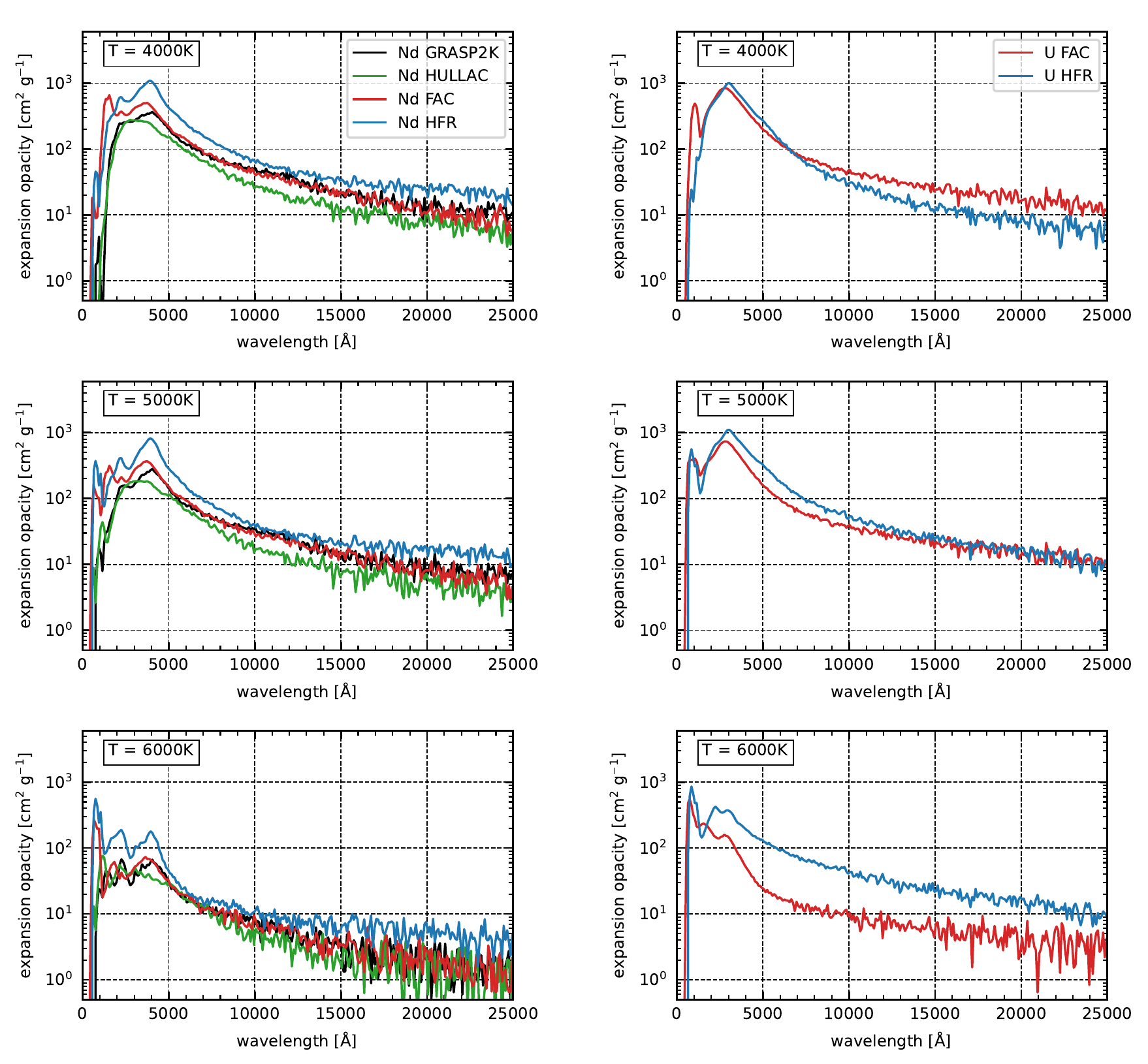}
    \caption{\textit{Left panels:} Expansion opacity of Nd at $T=4000$\,K, $5000$\,K and $6000$\,K and for $\rho=10^{-13}$\,g\,cm$^{-3}$ and $t=1$\,day with a bin width of 100\AA. Shown is the atomic data computed with the \FAC\, (red) and \HFR\, (blue) codes as well as published atomic data using \GRASP\, \citep[black, ][]{2019ApJS..240...29G} and \HULLAC\, \citep[green, ][]{2020MNRAS.496.1369T}. In all panels we show the \texttt{22 config + extra CI} \FAC\, and the \texttt{n=9} \HFR\, calculations (see Tables~\ref{tab:FAC_configs} and \ref{tab:HFR_configs}). The bin width was chosen such that variations between bins in the NIR are smaller than the differences between codes. The choice of the bin width does not affect the absolute scale of the expansion opacities but only acts as a smoothing parameter. \textit{Right panels:} Same as in left panels but for U instead of Nd.}
    \label{fig:expansion_opacity_Nd_U_all}
\end{figure*}

The opacity depends on the physical conditions of the ejecta: temperature, density and time after coalescence. To facilitate comparisons with other studies we adopt typical parameters for the ejecta at 1\,day after the merger. We choose $T=5000$\,K in accordance with the continuum of AT2017gfo inferred from the spectrum at 1.4\,days, as well as a density of $\rho=10^{-13}$\,g\,cm$^{-3}$, characteristic for an ejecta mass of $\sim 10^{-2}$\,$M_\odot$ distributed uniformly within a sphere expanding at $0.1c$. The choice of temperature is motivated by the fact that in LTE the radiation temperature equals the plasma temperature. Additionally, in Fig.~\ref{fig:expansion_opacity_Nd_U_all}, we present expansion opacities for temperatures of $T=4000$\,K and $T=6000$\,K, representative for a fully singly or doubly ionised plasma, respectively.

\begin{figure}
    \includegraphics[width=\columnwidth]{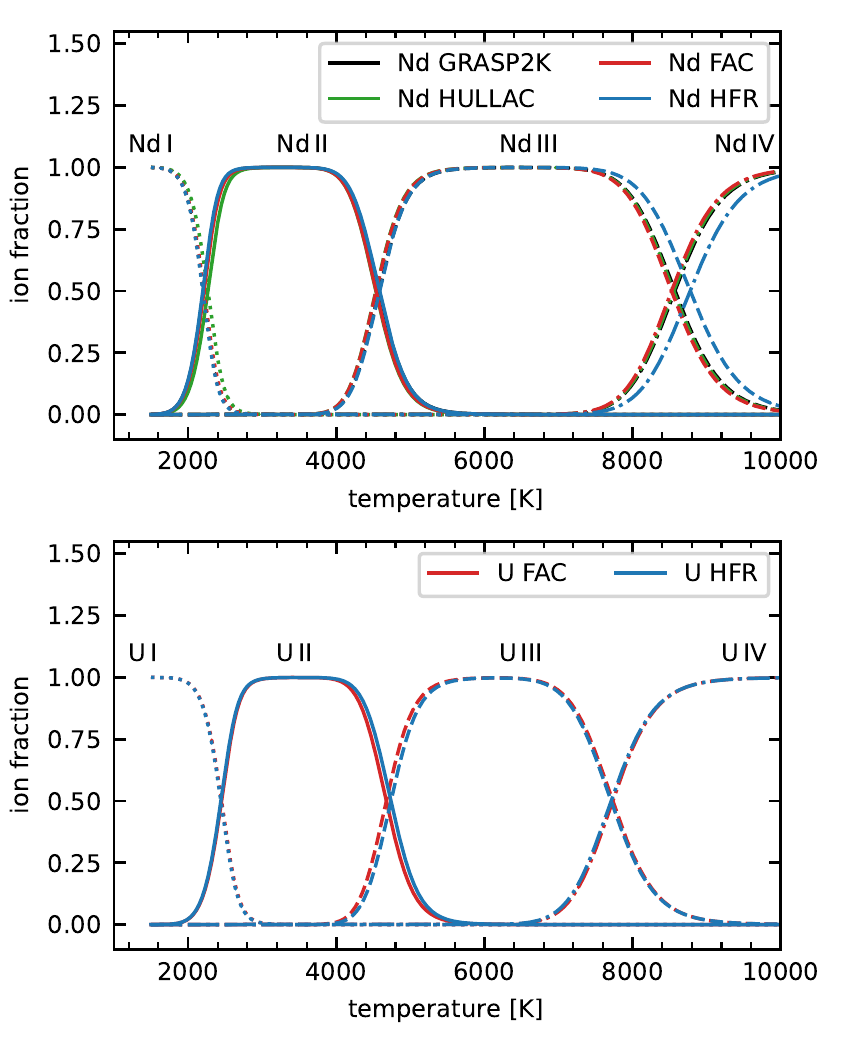}
    \caption{Top panel: Ion fractions of neutral to triply ionised Nd as a function of temperature for the atomic data computed with the \FAC\, (red) and \HFR\, (blue) codes as well as literature data computed with \GRASP\, \citep[black, ][]{2019ApJS..240...29G} and \HULLAC\, \citep[green, ][]{2020MNRAS.496.1369T}. Bottom panel: Same as above but for U.}
    \label{fig:ion_balance_Nd_U}
\end{figure}

\begin{figure}
    \includegraphics[width=\columnwidth]{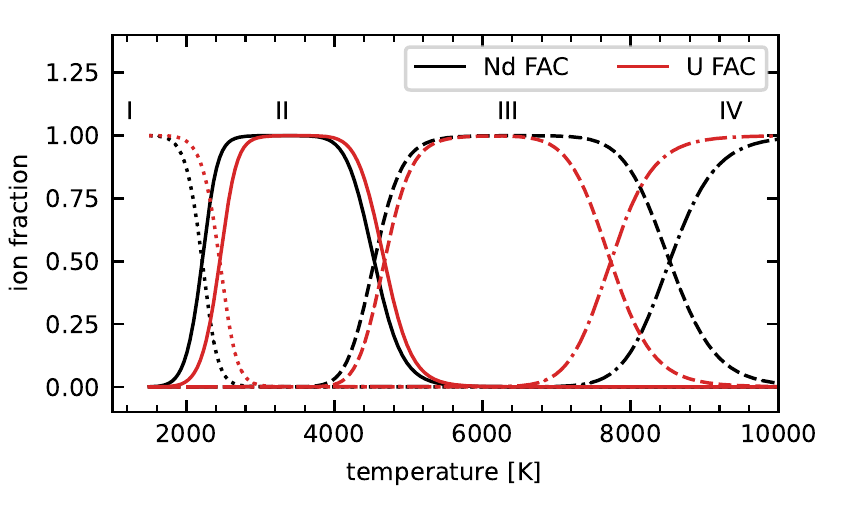}
    \caption{Comparison of the ion fractions of neutral to triply ionised Nd and U as computed with \FAC. Even though differences between the onset of ionisation stages of Nd and U appear to be small, due to the large gradient in the transition regions the opacity will be affected.}
    \label{fig:Nd_U_ion_fraction}
\end{figure}

In this study, we limit ourselves to LTE conditions only \citep[see ][for a discussion of non-LTE effects on kilonova opacities]{2022MNRAS.513.5174P, 2021MNRAS.506.5863H}. For the above-stated ejecta properties we present the resulting ionisation balance as a function of temperature in Fig.~\ref{fig:ion_balance_Nd_U} for the atomic data computed with \FAC\,and \HFR, as well as published data using the \GRASP\,and \HULLAC\,atomic structure codes \footnote{We note that our LTE ionisation balance calculations deviate from those presented in \citet[][]{2019ApJS..240...29G} and \citet{2020MNRAS.496.1369T}. While in this study all atomic levels are used to compute the partition functions, the studies as mentioned earlier use only the ground state (M. Tanaka, priv. communication). At the temperatures prevalent in kilonova ejecta after $\approx1$\,day, this can lead to a factor of order unity in the ratio of the partition functions entering the Saha equations compared to neglecting the temperature dependence through the Boltzmann factor (For \ion{Nd}{ii} at 5000\,K, the ratio of the partition functions $Z_{\ion{Nd}{ii}} / Z_{\ion{Nd}{iii}}$ is  $154.26\,/\,39.44\,=\,3.91$, while taking only the ground states $Z_{\ion{Nd}{ii},\mathrm{ground}} / Z_{\ion{Nd}{iii}, \mathrm{ground}}$ gives 8\,/\,9\,=\,0.89). As the ion density also enters the expression of the Sobolev optical depth, presented opacities are expected to differ between this work and the aforementioned studies. We emphasise that the different partition function treatment only affects the transition region between ionisation stages, for example, for II$\leftrightarrow$III, between 4000 and 5500\,K, as long as the Boltzmann level populations are properly normalised. All expansion opacities shown throughout this work, including those from published atomic data by \GRASP\, \citep[][]{2019ApJS..240...29G} and \HULLAC\, \citep[][]{2020MNRAS.496.1369T}, were computed using the full partition functions.}. 

Ionisation energies for all ions were taken from the NIST ASD \citep{NIST_ASD}. For ions other than singly or doubly charged, we used measured levels from NIST, so that the only changes between calculations are due to the atomic data of the singly and doubly charged ions. For temperatures in the range between 2000\,K and 10000\,K, typical for kilonova ejecta within the first week after merger, only levels up to a few eV can be populated and thus contribute to the partition function. We find that for the ions considered in this study, these low-lying levels are sufficiently well known such that the partition functions computed from measured (NIST ASD or SCASA) and calculated levels agree within $5\%$. 

For most highly charged ions only the ground state and the corresponding statistical weight $g$ are known. Unknown excited levels of highly charged ions ($\gtrapprox 5+$) have a negligible effect on the computed ionisation balance for temperatures between 2000 and 10000\,K. We find that for all sets of atomic data considered in this study, the ionisation balance is in good agreement for temperatures corresponding to the continuum of AT2017gfo within the first week ($\approx 2500$\,K to $8000$\,K). However, we note that for the often used temperature of $T=5000$\,K the balance for both Nd and U falls on the steep slope between singly and doubly ionised, where a small difference in the atomic data -- for example from the partition functions -- can have a significant effect on the ionisation balance and thus on the resulting opacity. 

Fig.~\ref{fig:Nd_U_ion_fraction} shows a comparison of the ionisation balance of Nd and U using the atomic data from our calibrated \verb|22 config + extra CI| \FAC\, calculations. We find that going from the lanthanide Nd to the actinide U shifts the transition from singly to doubly ionised states to slightly higher temperatures ($\Delta T\approx 130$\,K). As a result, when the temperature of the kilonova ejecta drops, \ion{Nd}{iii} will start to recombine shortly before \ion{U}{iii}. As we did not compute atomic data for neutral and triply ionised ions we cannot make any quantitative claims about the $\mathrm{I} \longleftrightarrow \mathrm{II}$ and $\mathrm{III} \longleftrightarrow \mathrm{IV}$ ionisation transitions.

\subsubsection{Expansion Opacity}
\label{sec:expansion_opacity}

Following Equations \ref{equ:expansion_opacity} to \ref{equ:boltzmann_excitation}, we compute expansion opacities for Nd and U (Fig.~\ref{fig:expansion_opacity_Nd_U_all}). To emphasise the effect of the temperature on the computed opacity we show the expansion opacity for 4000\,K, 5000\,K and 6000\,K, corresponding to the fully singly ionised, partially singly and doubly ionised and fully doubly ionised cases, respectively 
(see ion fractions of Fig.~\ref{fig:ion_balance_Nd_U}). 

We find that the atomic data calculated by \GRASP\ and \HULLAC\ as well as by our \FAC\ models typically yield lower opacities (especially in the shorter UV wavelength range) than the atomic data from our \HFR\ models. This is evident from the level structure, where close-to-ground states in the former have higher excitation energies than the ones in the latter, leading to a reduced opacity due to the exponential Boltzmann factor $\exp(-E_k/k_B T)$. Additional configurations included in the HFR models and, as discussed in Sections \ref{sec:Nd_atomic_data} and \ref{sec:U_atomic_data}, the different optimization procedures used in the \FAC\ , \GRASP\ , \HULLAC\ and \HFR\ codes (which are based on a restricted number of configurations in the first three and on all the configurations included in the model in the last one) could explain such differences.

\begin{figure*}
    \includegraphics[width=\textwidth]{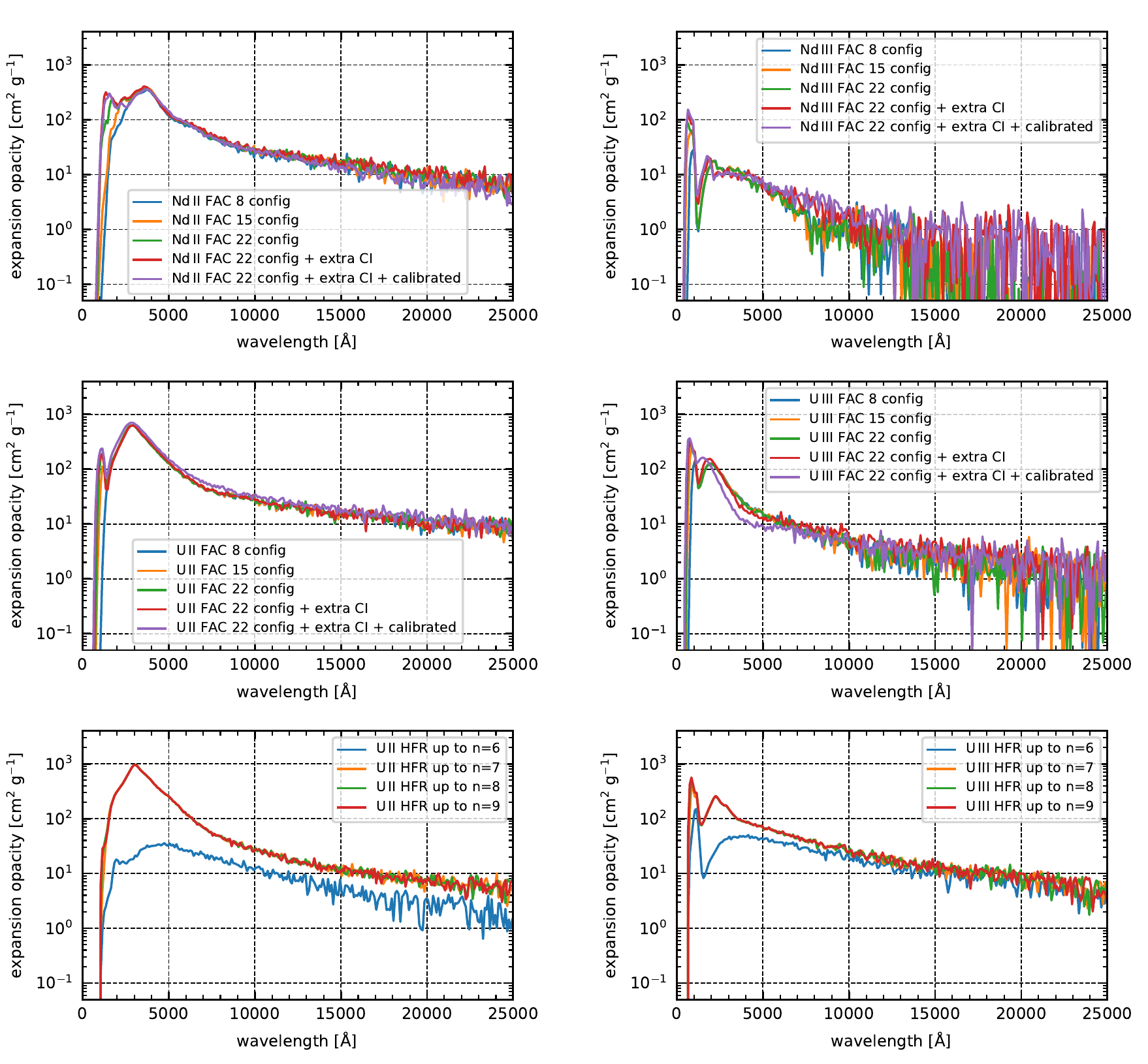}
    \caption{\textit{Top panels:} Expansion opacity of \ion{Nd}{ii} (left) and \ion{Nd}{iii} (right) at $T=5000$\,K, $\rho=10^{-13}$\,g\,cm$^{-3}$ and $t=1$\,day with a bin width of 100\AA~for \FAC\, calculations with varying numbers of included configurations (see Table~\ref{tab:FAC_configs}). \textit{Middle and bottom panels:} Same as top panels but for U instead of Nd (middle panels) and for \HFR\, instead of \FAC\, (bottom panels).}
    \label{fig:convergence_expansion_opacity_FAC_HFR}
\end{figure*}

Peak expansion opacities of Nd, for which published atomic data is available, differ by about 0.5\,dex among the calculations considered. While there is some variation at longer wavelengths due to differences in the level density near the ground state, the total number of configurations included in the atomic structure calculations particularly affects the opacity red-wards of the ionisation edge at shorter wavelengths of about 1200~\.Angstroms. In this region the sharp drop in opacity occurs at longer wavelengths for the GRASP2K and HULLAC curves (see Fig.~\ref{fig:line_binned_opacity} for a clearer view of this portion of the opacity spectrum). This is due to highly energetic states not present in smaller calculations, which cannot be populated in LTE, but serve as upper levels for transitions from the populated near-ground levels. These transitions only contribute opacity for photon energies $E_u-E_l \approx E_\mathrm{ion}$. Fig.~\ref{fig:convergence_expansion_opacity_FAC_HFR} illustrates the convergence of the opacity with an increasing number of configurations treated in our \FAC\, and \HFR\, atomic structure calculations. While the opacity at the long-wavelength tail remains virtually unchanged for all ions, with only small variations from a small shift of the involved energy levels occurring, additional opacity components come into play at short wavelengths. The location and magnitude of these components depend on the ions and calculation strategies involved. 

For the Nd atomic data computed with \FAC\, we notice an increase in the opacity with increasing number of included configurations between 3000\,\AA\,($\approx 4$\,eV) and 1100\,\AA\,($\approx 11.5$\,eV) and between 2000\,\AA\,($\approx 6$\,eV) and 650\,\AA\,($\approx 19$\,eV), respectively, for \ion{Nd}{ii} and \ion{Nd}{iii}. The lower bounds are close to the ionisation energy for both ions (11.6\,eV, and 19.8\,eV, respectively). Increasing the number of included configurations beyond 22 (\ion{Nd}{ii}) and 15 (\ion{Nd}{iii}) only had a minor effect on the opacity. In either case, if applied to LTE kilonova radiative transfer modelling, including the 15 lowest configurations should suffice for capturing the majority of the opacity, as the radiation field even in the early kilonova evolution ($\approx$1\,day) is negligible at wavelengths shorter than about 2000\,\AA. A 10000\,K blackbody, as observed in the earliest (0.5\,day) spectrum of AT2017gfo \citep[][]{2022MNRAS.515..631G}, contains only about $3\%$ of its flux at wavelengths shorter than 3000\,\AA. 

Compared to Nd the \FAC\,calculations of \ion{U}{ii} and \ion{U}{iii} require less configurations until convergence is achieved. Going beyond 15 configurations affects the opacity only marginally below 2000\,\AA\,or for photon energies above $\approx 6$\,eV. We note, however, that the calibration to measured levels (purple curves in Fig.~\ref{fig:convergence_expansion_opacity_FAC_HFR}) has a much greater effect on the U opacity than the Nd opacity. 

Our \HFR\,calculations follow a different strategy, in which additional configurations are added based on the maximum shell number $n$ of the excited states. From Fig.~\ref{fig:convergence_expansion_opacity_FAC_HFR} it is apparent that the inclusion of configurations with $n \leq 7$ is sufficient to capture the bulk of the atomic opacity, both around the peak and the tail. Transitions from near-ground-state levels to the $n=7$ shell dominate the opacity across the full optical and NIR wavelength range. Near the peak (1000\,\AA\,to 4000\,\AA),the opacity is almost unaffected by the inclusion of transitions to levels from the $n>7$ configurations.

The opacity tail at long wavelengths is a direct measure of the density of levels near the ground state. Again, as only levels close to the ground state can be populated at temperatures of a few $10^3$\,K, the upper level needs to be within $\approx 1$\,eV (corresponding to $\lambda\approx$ 1\,$\mu$m) to the lower level. Similar to the peak expansion opacities, we find significant differences between structure calculations. However, the opacities from the calibrated \verb|22 config + extra CI | \FAC\, and the \GRASP\, calculations agree exceptionally well both near the peak and on the tail, for both \ion{Nd}{ii} and \ion{Nd}{iii}. 

\begin{figure}
    \includegraphics[width=\columnwidth]{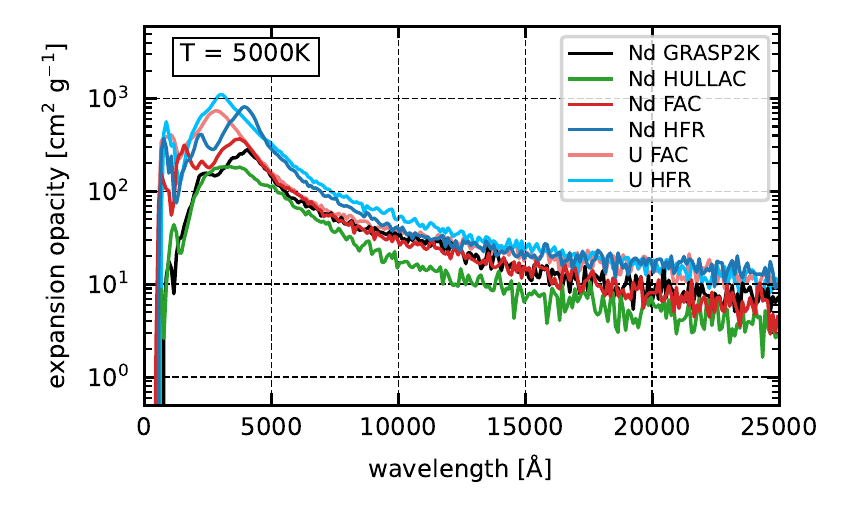}
    \caption{Comparison of the expansion opacities of Nd and U at $T=5000$\,K, $\rho=10^{-13}$\,g\,cm$^{-3}$ and $t=1$\,day computed with the \FAC\, (red) and \HFR\, (blue) codes and from published data from \GRASP\, \citep[black, ][]{2019ApJS..240...29G} and \HULLAC\, \citep[green, ][]{2020MNRAS.496.1369T}.}
    \label{fig:expansion_opacity_Nd_U_5000K}
\end{figure}

Next, we explore whether actinides opacities are comparable to lanthanides, or possibly, even higher. Fig.~\ref{fig:expansion_opacity_Nd_U_5000K} shows the expansion opacities of Nd and U computed at 5000\,K. We find peak expansion opacities of $\kappa_\mathrm{exp}^{\mathrm{Nd, \texttt{FAC}}}=368$\,cm$^{2}$~ g$^{-1}$ and $\kappa_\mathrm{exp}^{\mathrm{U}, \texttt{FAC}}=736$\,cm$^{2}$~g$^{-1}$ using \FAC\,(an increase by a factor of 2) and $\kappa_\mathrm{exp}^{\mathrm{Nd}, \texttt{HFR}}=815$\,cm$^{2}$~g$^{-1}$ and $\kappa_\mathrm{exp}^{\mathrm{U}, \texttt{HFR}}=1103$\,cm$^{2}$~g$^{-1}$ (an increase by a factor of 1.35) for \HFR. A similar behaviour can be seen in the atomic data from \citet[][]{2020MNRAS.493.4143F, 2023MNRAS.519.2862F} shown in Fig.~\ref{fig:line_binned_opacity}. At temperatures of 4641\,K (0.4\,eV) the opacity of U is higher than that of Nd by about a factor of 4. While there is significant variation in the opacities of any given ion from different structure codes, the resulting opacity of U for a given code always seems to be significantly higher than that of Nd.

\subsubsection{Line-binned opacities}
\label{sec:line_binned}

Another frequency-dependent opacity formalism is the so-called line-binned opacity \citep[][]{2020MNRAS.493.4143F, 2023MNRAS.519.2862F}, 
\begin{equation}
    \kappa^{\rm bin}_{\nu} = \frac{1}{\Delta \nu}\frac{\pi e^2}{\rho m_e c}\sum_{l} N_l\, |f_{l}| \,,
    \label{equ:line_binned_opacity}
\end{equation}
where $\rho$ is the ejecta density, $\Delta \nu$ are the widths of the frequency bins containing lines $l$ with number densities of the lower level $N_l$ and oscillator strengths $f_l$. This expression is obtained by replacing the line profile with a flat distribution across the corresponding bin \citep[][]{2020MNRAS.493.4143F}. Compared with the expansion opacity (Equation~\ref{equ:expansion_opacity}), it can be pre-computed as Equation~\ref{equ:line_binned_opacity} is independent of the expansion time $t$. 

\begin{figure*}
    \includegraphics[width=\textwidth]{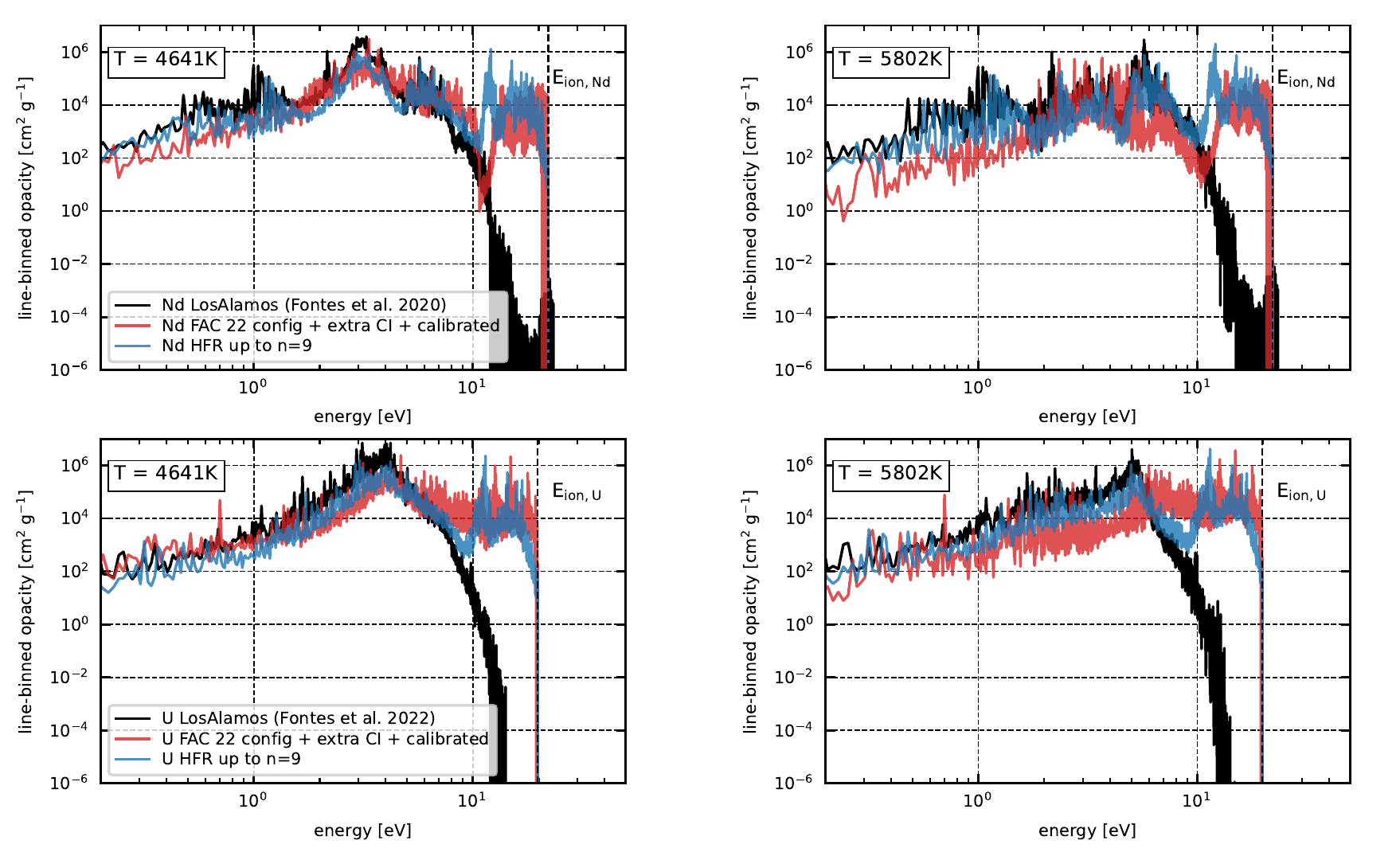}
    \caption{Line-binned opacities of Nd (upper panels) and U (lower panels) at $\rho=10^{-13}$\,g\,cm$^{-3}$, $t=1$\,day and $T=4841$\,K ($0.4$\,eV, left panels) and $T=5802$\,K ($0.5$\,eV, right panels), using re-binned published opacities from the \texttt{Los Alamos suite of atomic physics and plasma modeling codes} \citep[black, ][]{2020MNRAS.493.4143F, 2023MNRAS.519.2862F}, the \FAC\,22 config + extra CI calibrated and the \HFR\, up to $n=9$ calculations.}
    \label{fig:line_binned_opacity}
\end{figure*}

Fig.~\ref{fig:line_binned_opacity} shows the Nd and U line binned opacities from \citet[][]{2020MNRAS.493.4143F, 2023MNRAS.519.2862F} as well as the largest calibrated \FAC\, and \HFR\, calculations for temperatures equivalent to $0.4\,$eV and $0.5\,$eV (4641\,K and 5802\,K, respectively). For low photon energies (long wavelengths) we find good agreement between our calculations and the ones from \citet{2020MNRAS.493.4143F, 2023MNRAS.519.2862F}. In the long wavelength regime, our \HFR\, opacities come closer to those of \citet{2020MNRAS.493.4143F, 2023MNRAS.519.2862F} as they exhibit a higher density of levels near the ground state (lower level energies than what is reported on the NIST ASD). However, for high photon energies the published line-binned opacities from \citep[][]{2020MNRAS.493.4143F} display a sharp cut-off at $\approx 10$\,eV, which is not seen in the \FAC\,and \HFR\,data. We identify the additional high-energy opacity component as transitions from near-ground states to highly excited states close to the ionisation edge (see Section~\ref{sec:expansion_opacity} for a similar effect in the case of expansion opacities) from configurations not included in the calculations by \citet[][]{2020MNRAS.493.4143F, 2023MNRAS.519.2862F}. In the high-temperature case, in which neodymium and uranium are doubly ionised, we find variations of about one order of magnitude between the various calculations. This is likely a result of the difficult calibration for \ion{Nd}{iii} and \ion{U}{iii} and the resulting uncertainty in the density of levels near the ground state (see discussion in Section~\ref{sec:Nd_atomic_data}).

\subsubsection{Planck Mean Opacity}

\begin{figure}
    \includegraphics[width=\columnwidth]{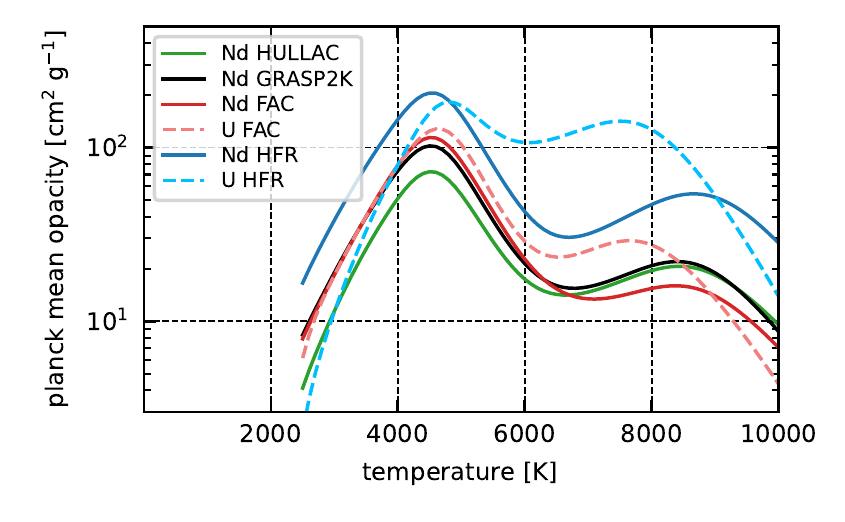}
    \caption{Comparison of Planck's mean opacity as a function of temperature and for $\rho=10^{-13}$\,g\,cm$^{-3}$ and $t=1$\,day for Nd and U computed with the \FAC\, (red colors) and \HFR\, (blue colors) codes and from published data from \GRASP\, \citep[black, ][]{2019ApJS..240...29G} and \HULLAC\, \citep[green, ][]{2020MNRAS.496.1369T}. }
    \label{fig:plack_mean_opacity_Nd_U}
\end{figure}

In addition to the wavelength dependent expansion opacity we also compute wavelength-independent 'grey' opacities in the form of the Planck mean opacity defined as 
\begin{equation}
    \kappa_{\mathrm{mean}} = \frac{\int_0^\infty B_\lambda(T)\kappa_\mathrm{exp}(\lambda)d\lambda}{\int_0^\infty B_\lambda(T)d\lambda} .
\end{equation}
Grey opacities are predominantly used for light curve modelling, in which the diffusion of photons due to opacity is the dominant effect. Use of the Planck mean opacity requires the radiation field to be a blackbody spectrum, which, in the case of kilonovae, is well justified during its early evolution \citep[see blackbody fits in ][]{2021FrASS...7..108P, 2017Natur.551...75S, 2019Natur.574..497W, 2022MNRAS.515..631G}. At the point at which transitions begin to dominate the spectrum -- in AT2017gfo this happened $\approx 5$\,days after merger -- the usefulness of the Planck mean opacity begins to break down. A comparison of the Planck mean opacities for atomic data computed in this work and from other sources \citep[][]{2019ApJS..240...29G, 2020MNRAS.496.1369T} is shown in Fig.~\ref{fig:plack_mean_opacity_Nd_U}. The peak of the Planck mean opacity is located for all ions considered at $T\approx4500$\,K. The peak location follows from the convolution of the blackbody spectrum with the frequency-dependent expansion opacity (Fig.~\ref{fig:expansion_opacity_Nd_U_all}). A lower temperature leads to an increased abundance of singly ionised ions which have a much larger opacity than their doubly ionised counterparts, but a blackbody that peaks farther in the red, where the opacity is decreasing with $\lambda^{-1}$ \citep[][]{2022Atoms..10...18S}. Even though the expansion opacity does not change significantly between 2500\,K and 4500\,K, the reduced overlap between the Planck spectrum and the wavelength-dependent opacity leads to a sharp drop of the Planck mean opacity below 4000\,K. Above 5500\,K the Planck mean opacity remains almost constant, as the doubly ionised state dominates the ionisation balance and the blackbody spectrum peaks in the optical (5000\,\AA\,to 10000\,\AA\,for temperatures at which doubly ionised Nd/U prevails). The local maximum near 8000\,K followed by a sharp drop is due to the III$\leftrightarrow$IV transition near that temperature, surpassing the increasing overlap of the Planck spectrum with the peak of the opacity.

\begin{table}
    \centering
    \caption{Planck mean opacity for a pure Nd or U plasma with $\rho=10^{-13}$ g\,cm$^{-3}$ and $t = 1$\,day after the merger. $\overline{\kappa}$ is the Planck mean opacity averaged over the temperature range T = 2500 -- 7500\,K and T = 5000 -- 10000\,K, respectively.}
    \label{tab:planck_mean_opacities}
    \begin{tabular}{lccc}
        \toprule
        Ion & $\overline{\kappa} [$cm$^2 $g$^{-1}$] & $\overline{\kappa} [$cm$^2 $g$^{-1}$] & peak $\kappa [$cm$^2 $g$^{-1}$]\\
        & 2500 -- 7500\,K & 5000 -- 10000\,K\\
        \toprule
        Nd \GRASP\, &41.29 & 21.63 &  102.39\\
        Nd \HULLAC\, & 29.95 & 18.98 & 72.66\\
        Nd \FAC\, & 44.10 & 19.79 & 114.35\\
        Nd \HFR\, & 83.27 & 48.86 & 206.32\\
        U \FAC\, & 51.20 & 26.84 & 128.17\\
        U \HFR\, & 99.27 & 99.11 & 183.22\\
        \midrule
        \bottomrule
    \end{tabular}
\end{table}

We find mean variations in the Planck mean opacity between 2500\,K and 10000\,K of  $\overline{\kappa_\mathrm{mean}^{\mathrm{Nd,  \texttt{HFR}}}/\kappa_\mathrm{mean}^{\mathrm{Nd,  \texttt{FAC}}}}=2.50$, $\overline{\kappa_\mathrm{mean}^{\mathrm{Nd,  \texttt{HULLAC}}}/\kappa_\mathrm{mean}^{\mathrm{Nd,  \texttt{FAC}}}}=0.94$ and $\overline{\kappa_\mathrm{mean}^{\mathrm{Nd,  \texttt{GRASP2K}}}/\kappa_\mathrm{mean}^{\mathrm{Nd,  \texttt{FAC}}}}=1.11$. Similarly, we find $\overline{\kappa_\mathrm{mean}^{\mathrm{U,  \texttt{HFR}}}/\kappa_\mathrm{mean}^{\mathrm{U,  \texttt{FAC}}}}=2.96$. A summary of the peak and temperature averaged Planck mean opacities is given in Tab.~\ref{tab:planck_mean_opacities}.

\section{Conclusions}
\label{sec:conclusions}

We computed energy levels and oscillator strengths for the singly and doubly ionised ions of neodymium and uranium, using the \FAC\,and \HFR\, atomic structure codes. In the calculations we included a maximum of 30 (\FAC) and 27 (\HFR) non-relativistic electronic configurations.

To obtain better agreement between computed and experimental level energies from the NIST ASD \citep[][]{NIST_ASD} and SCASA \citep{SCASA} we optimised the mean fictitious configuration used in the construction of the local central potential in the \FAC\,calculations. Employing this optimisation reduces the mean deviation from NIST data among the lowest 100 levels by a factor of two. An automatised methodology of optimisation is being tested at the time of writing this paper, which will be used to try to provide a more reliable, while still complete, set of atomic data for relevant lanthanide and actinide elements. (F. Silva 2023, In prep.). In addition to the potential optimisation, we calibrated the calculated level energies, split into groups of $J$-$P$, to experimental data. 

We also investigated the convergence behaviour of our atomic data with the number of included electronic configurations. We confirm that $\approx15$ configurations up to the $n=7$ (Nd) or $n=8$ shell (U) are sufficient to capture most the opacity that falls into the optical and NIR wavelength regions. The inclusion of additional configurations beyond those only yields minor contributions to the opacity close to the ionisation potential of the ions. 

For each of the ions we computed bound-bound E1 transitions. From the calculated transitions we derive wavelength-dependent opacities assuming LTE conditions in the plasma. Included electron configurations were chosen such that internal convergence of the atomic opacities is obtained. We find good agreement between the optimised and calibrated \FAC\,atomic data and calculations of Nd from \citet[][]{2019ApJS..240...29G}, \citet[][]{2020MNRAS.496.1369T} and \citet[][]{2020MNRAS.493.4143F}. Our ab-initio calculations using the \HFR\, code yield higher opacities than the corresponding \FAC\, calculations. This larger opacity is a direct consequence of the higher density of levels near the ground state in the calculated \HFR \, atomic data. Doubly ionised ions, for which less experimental data is available, display a higher opacity variance between codes than singly ionised ions. We identify the density of levels near the ground state as the main reason for the opacity difference between codes, while the total number of included levels and lines only has a minor effect on the opacity. The higher level density predicted by \HFR\ is partially due to the additional configurations included in the \HFR\ computations, but such extra configurations often give rise to levels, which, for typical temperatures, cannot be populated in LTE. Only transitions between these and near ground-state levels contribute to the opacity, predominantly at wavelengths close to the excitation energy of the upper level ($\approx$ 1000\,\AA\ -- 3000\,\AA). However, the optimization procedures in \HFR\ and \FAC\ are fundamentally different, insofar as the average energies of all the configurations within the model are minimised in the former, while the potential optimization is based on a restricted number of configurations in the latter (as well as in the \GRASP\ and \HULLAC\ codes that were respectively used in \citet[][]{2019ApJS..240...29G} and \citet[][]{2020MNRAS.496.1369T}). As a consequence, this difference can explain the lower energies of the predicted \HFR\ levels, thus the higher density of levels near the ground state, leading to higher opacities.

In this study, we confirm that actinides, such as uranium, have an opacity that is a factor of a few higher than the opacity of the corresponding lanthanides, with ratios of 1.35 (\HFR), 2 (\FAC) and 4 \citep[\texttt{Los Alamos suite of atomic physics and plasma modeling codes},][]{2020MNRAS.493.4143F, 2023MNRAS.519.2862F} depending on the atomic structure code used. This result is of particular interest for the modelling of kilonovae, as the presence of even small amounts of actinides formed in the lowest-$Y_e$ ejecta can have strong implications on the derived synthetic light curves and spectra.

\section*{Acknowledgements}

AF, GL, LJS and GMP acknowledge support by the European Research Council (ERC) under the European Union’s Horizon 2020 research and innovation program (ERC Advanced Grant KILO-
NOVA No. 885281) and the State of Hesse within the Cluster Project ELEMENTS. JD and SG acknowledge financial support from F.R.S.-FNRS (Belgium). RFS acknowledges the support from National funding by FCT (Portugal), through the individual research grant 2022.10009.BD. RFS, JMS, JPM and PA acknowledge the support from FCT (Portugal) through research center fundings UIDP/50007/2020 (LIP) and UID/04559/2020 (LIBPhys), and through project funding 2022.06730.PTDC, "Atomic inputs for kilonovae modeling (ATOMIK)". PQ is Research Director of the F.R.S.-FNRS. HCG is a holder of a FRIA fellowship. PP is a Research Associate of the Belgian Fund for Scientific Research F.R.S.-FNRS. 

This work has been supported by the Fonds de la Recherche Scientifique (FNRS, Belgium) and the Research Foundation Flanders (FWO, Belgium) under the Excellence of Science (EOS) Project (number O022818F and O000422F). Computational resources have been provided by the Consortium des Equipements de Calcul Intensif (CECI), funded by the F.R.S.-FNRS under Grant No. 2.5020.11 and by the Walloon Region of Belgium.

We thank G.~Gaigalas, M.~Tanaka, D.~Kato and C.~J.~Fontes for making some or all of their calculated atomic data public, facilitating benchmarks with other codes.

This research made use of Astropy, a community-developed core Python package for Astronomy \citep{2013A&A...558A..33A, 2018AJ....156..123A}, as well as numpy \citep{2011CSE....13b..22V}, scipy \citep{scipy}, pandas \citep{mckinney-proc-scipy-2010} and matplotlib \citep{2007CSE.....9...90H}.
\section*{Data Availability}
 
The data underlying this article is available from the authors upon reasonable request.



\bibliographystyle{mnras}
\bibliography{references} 

\bsp	
\label{lastpage}
\end{document}